\documentclass[12pt]{iopart}

\usepackage{iopams}
\usepackage{url}
\usepackage{hyperref}
\usepackage{stackrel}
\usepackage{mathabx}
\usepackage{subcaption}
\usepackage{graphicx}
\usepackage{xcolor}
\usepackage{sidecap}
\usepackage{cite}
\usepackage{braket}
\usepackage{pdflscape}
\usepackage{tikz}

\newcommand{\tfrac}[2]{\textstyle\frac{#1}{#2}}
\newcommand{\dfrac}[2]{\displaystyle\frac{#1}{#2}}

\hypersetup{
    unicode=false,
    pdftoolbar=true,
    pdfmenubar=true,
    pdffitwindow=false,
    pdfstartview={FitH},
    pdftitle={Information-Theoretic and Operational Measures of Quantum Contextuality},
    pdfauthor={A C Gunhan and Z Gedik},
    pdfsubject={Quantum Contextuality},
    pdfkeywords={quantum contextuality, Kochen-Specker theorem, mutual information energy, uncertainty relations, KCBS inequality, Majorana stellar representation},
    pdfnewwindow=true,
    colorlinks=true,
    linkcolor=magenta,
    citecolor=blue,
    filecolor=magenta,
    urlcolor=cyan
}

\begin{document}
\title{Information-Theoretic and Operational Measures of Quantum Contextuality}

\author{A C G\"unhan$^{1,\ast}$ and Z Gedik$^2$}
\address{$^1$ Department of Physics, Mersin University, \c{C}iftlikk\"oy Merkez Yerle\c{s}kesi, Yeni\c{s}ehir, Mersin, 33150 T\"urkiye}
\address{$^2$ Faculty of Engineering and Natural Sciences, Sabanci University, Tuzla, Istanbul, 34956 T\"urkiye}

\ead{\href{mailto:alicangunhan@mersin.edu.tr}{alicangunhan@mersin.edu.tr}} 
\ead{\href{mailto:gedik@sabanciuniv.edu}{gedik@sabanciuniv.edu}}

\begin{indented}
\item[]$^\ast$ Corresponding author.
\end{indented}

\begin{abstract}
We propose an information--theoretic framework for quantifying Kochen--Specker contextuality. Two complementary measures are introduced: the mutual information energy, a state--independent quantity inspired by Onicescu's information energy that captures the geometric overlap between joint eigenspaces within a context; and an operational measure based on commutator expectation values that reflects contextual behavior at the level of measurement outcomes. We establish a hierarchy of bounds connecting these measures to the Robertson uncertainty relation, including spectral, purity--corrected, and operator norm estimates. The framework is applied to the Klyachko--Can--Binicio\u{g}lu--Shumovsky (KCBS) scenario for spin-1 systems, where all quantities admit closed--form expressions. The Majorana--stellar representation furnishes a common geometric platform on which both the operational measure and the uncertainty products can be analyzed. For spin-1, this representation yields a three-dimensional Euclidean-like visualization of the Hilbert space in which, states lying on a plane exhibit maximum uncertainty for the observable along the perpendicular direction; simultaneous optimization across all KCBS contexts singles out a unique state on the symmetry axis. Notably, states achieving the optimal sum of uncertainty products exhibit vanishing operational contextuality, while states with substantial operational contextuality satisfy a nontrivial Robertson bound---the two extremes are achieved by distinct quantum states.
\end{abstract}


\maketitle

\section{Introduction: Contextuality and Measurement in Quantum Theory}
\label{sec:Intro}

The process and outcome of measurement are fundamentally different in classical and quantum mechanics.  
In classical (Newtonian) mechanics, the result of a measurement is assumed to exist prior to the act of measurement and remains unaffected by it—only one such measurement can be performed at a time.  
At the microscopic scale, however, this is no longer the case.  
Quantum theory allows an arbitrary number of observables to be measured simultaneously on a single system\footnote{Here, “simultaneous measurement’’ simply means that the outcomes of different observables can be obtained within the same measurement procedure. Mathematically, a single generalized measurement (POVM, Positive Operator-Valued Measure) may contain very many—indeed, even continuously many—outcomes. In this sense, one quantum measurement can encode information about infinitely many observables, even though not all of them can have sharp values at the same time.}, a feature often referred to as \textit{quantum parallelism}~\cite{Nielsen_Chuang_2010}.  
As a result, the state of a quantum system cannot, in general, be prepared in a dispersion-free manner~\cite{Bell1966}, nor can definite outcomes be ascribed to measurements prior to observation~\cite{Zeilinger2005}.  

If a system is in an eigenstate of one observable, it is necessarily not an eigenstate of another that does not commute with it—even though both can be measured jointly.  
Therefore, the outcome of measuring an observable cannot, in general, be independent of the choice of other compatible measurements performed concurrently~\cite{Hermann1935a, Hermann1935b, Peres1990}.  
Formally, for observables $A$, $B$, and $C$ represented by Hermitian operators $\hat{A}$, $\hat{B}$, and $\hat{C}$ acting on a Hilbert space $\mathcal{H}$ of dimension $d \ge 3$, which satisfy
\begin{equation}\label{eq:context}
[\hat{A}, \hat{B}] = 0, \qquad
[\hat{B}, \hat{C}] = 0, \qquad
[\hat{A}, \hat{C}] \ne 0,
\end{equation}
the outcome of a measurement of $B$ depends on whether it is performed alone, together with $A$, or together with $C$~\cite{Gleason1957, Specker1960}.  
In such a situation, the information obtained by measuring $(A,B)$ differs from that obtained by measuring $(C,B)$; hence the measurement of $B$ is \textit{contextual} within the set $\{A,B,C\}$~\cite{KochenSpecker1967, Redhead1987, Peres1991, CabelloRev22}.  
Accordingly, the question  
\textit{``How does the information obtained from the measurement of $(A,B)$ differ from that of $(C,B)$?''}  
naturally quantifies the \textit{amount of contextuality} associated with $B$ in the context $\{A,B,C\}$.

Several quantitative measures of contextuality have been proposed in the literature, most of which are formulated within probabilistic or operational frameworks.  
Svozil quantified contextuality in terms of the ``frequency of contextual assignments'' in a forced tabulation of truth values~\cite{Svozil2012}.  
Kleinmann et al. defined it through the \textit{memory cost} of measurement sequences, where the memory corresponds to the number of internal system states reached during sequential measurements, and the cost is its minimum value~\cite{Kleinmann_2011}.  
Grudka et al. introduced two measures: the \textit{cost of contextuality}, by analogy with the nonlocality cost~\cite{BrunnerCavalcanti2011, PopescuRohrlich1994}, and the \textit{relative entropy of contextuality}, analogous to the relative entropy of nonlocality~\cite{DamGillGrunwald2005}, showing that the latter is equivalent to a communication-based measure they also defined.  
Abramsky et al. proposed the \textit{contextual fraction}, quantifying how far a quasi-probability representation (allowing non-negative numbers summing to less than unity) can be from a true probability distribution~\cite{AbramskyBarbosaMansfield2017}.  
Kujala and Dzhafarov suggested three alternative measures~\cite{KujalaDzhafarov2019}: two as distances to the \textit{noncontextuality polytope}~\cite{Pitowsky1991, Kleinmann2012}—in which noncontextual systems are represented on or within the polytope surface—and a third based on quasi-probability distributions allowing negative values whose sum equals unity, where the magnitude of the negative part quantifies the degree of contextuality.

While these approaches differ in formulation, they all share the goal of quantifying the deviation from classical, noncontextual behavior.  
In the present work, we take an alternative route and formulate contextuality as an information–theoretic property, rooted in the geometric and algebraic relations between measurement subspaces.

The paper is organized as follows. In Section~\ref{sec:MIEdef}, we introduce the mutual information energy as a basis-independent, state-independent measure of contextuality based on the geometric overlap of joint eigenspaces. Section~\ref{sec:OperationalMeasure} develops the complementary operational measure, which captures contextual behavior at the level of measurement outcomes for a given quantum state, and establishes spectral, purity-corrected, and hybrid bounds that relate this operational quantity to the mutual information energy. In Section~\ref{sec:App_Context_Uncert}, we connect these bounds to the Robertson uncertainty relation, revealing a hierarchy that ties together the commutator structure, uncertainty products, and the geometric MIE quantity. Section~\ref{sec:KCBS} illustrates the full framework in the canonical KCBS scenario for a spin-1 system, where all quantities can be computed in closed form and visualized geometrically. We conclude in Section~\ref{sec:Conclusion} with a summary and outlook.

\section{Information--Theoretic Measure of Contextuality}
\label{sec:MIEdef}

Since the measurement of observables involves obtaining information from a quantum system, and since contextuality concerns how this information depends on the chosen set of compatible measurements, it is natural to treat contextuality as an \emph{information--theoretic property}~\cite{Shannon1949Book,Renyi1961}.  
A useful scalar quantity in this framework is the \textit{information energy}, originally introduced by Onicescu~\cite{onicescu1966theorie}. 
For a discrete random variable $X$ taking values $x_1, x_2, \dots, x_n$ with corresponding probabilities $p_1, p_2, \dots, p_n$, 
the information energy is defined as
\begin{equation}\label{eq:InfEnDef}
    \mathcal{E}(X) = \sum_{i=1}^n p_i^2.
\end{equation}
This quantity serves as an inverse measure of uncertainty: 
it reaches its maximum for a deterministic (pure) distribution and its minimum for a uniform one.

Motivated by Onicescu’s notion of information energy as a quadratic measure of concentration, we extend this idea to the quantum domain. Extending this concept to the quantum domain, we define the \textit{mutual information energy} (MIE) as a  basis--independent quantity that captures the informational overlap between the eigenspaces of different observables. For observables $A$, $B$, $C$ represented by Hermitian operators $\hat{A}$, $\hat{B}$, $\hat{C}$, 
the MIE is expressed in a projection--based form as
\begin{equation}\label{eq:MIE_proj_def}
E(B;A,C) \coloneq \frac{1}{d}\sum_{i,j}
\mathrm{Tr}\!\left[\left(\hat P^{A,B}_i \hat P^{C,B}_j \right)^2 \right],
\qquad d=\dim(\mathcal{H}) \ge 3,
\end{equation}
where $\hat P^{A,B}_i$ and $\hat P^{C,B}_j$ are the projectors onto the joint eigenspaces of 
$(\hat{A},\hat{B})$ and $(\hat{C},\hat{B})$, respectively.  
Each projector satisfies
\begin{eqnarray}
\hat{A}\hat{B}\hat P^{A,B}_i &=& a_i b_i \hat P^{A,B}_i, \label{eq:ABproj}\\
\hat{C}\hat{B}\hat P^{C,B}_j &=& c_j b_j \hat P^{C,B}_j. \label{eq:CBproj}
\end{eqnarray}
Equation~(\ref{eq:MIE_proj_def}) thus quantifies the degree of informational overlap 
between the subspaces associated with $\hat{A}$ and $\hat{C}$ 
within each contextual partition defined by $\hat{B}$.

When all three observables commute, their projectors coincide, 
$\hat P^{A,B}_i = \hat P^{C,B}_i$, and
\begin{equation}\label{eq:Eequals1}
E(B;A,C) = 1,
\end{equation}
indicating a fully noncontextual situation in which measurement outcomes are jointly definable. 
At the opposite extreme, when the eigenspaces of $\hat{A}$ and $\hat{C}$ are mutually unbiased 
within each eigenspace of $\hat{B}$~\cite{Ivanovic1981, Wootters1989363, Bengtsson2012374}, 
the projectors satisfy
\begin{equation}\label{eq:MutUnbTr}
\mathrm{Tr}\!\left[(\hat P^{A,B}_i \hat P^{C,B}_j)^2\right] 
    = \frac{\mathrm{dim}(\hat P^{A,B}_i)\,\mathrm{dim}(\hat P^{C,B}_j)}{d^2},
\end{equation}
a result whose proof is provided in the Supplementary Information (SI), section A. In this case the mutual information energy attains its minimum value,
\begin{equation}\label{eq:EMin}
E(B;A,C)=\frac{1}{d}.
\end{equation}
This lower bound remains valid even in the presence of degeneracies, 
provided that the relevant eigenspaces are uniformly unbiased with respect to each other.

Hence, $E(B;A,C)$ defines a \textit{basis--independent, projection--based measure of contextuality}, 
robust to degeneracies and directly linked to the geometric overlap of measurement subspaces.  
In the nondegenerate case where all eigenspaces are one--dimensional, 
each projector reduces to a rank--1 operator, $\hat P^{A,B}_i = |ab_i\rangle\langle ab_i|$, 
and equation~(\ref{eq:MIE_proj_def}) simplifies to
\begin{equation}\label{eq:MIE_rank1}
E(B;A,C) = \frac{1}{d}\sum_{i,j=1}^d |\langle ab_i | cb_j \rangle|^4.
\end{equation}
This form explicitly shows that $E(B;A,C)$ depends solely on the relative geometry of the eigenspaces of $\hat{A}$ and $\hat{C}$—a purely quantum manifestation of informational compatibility within the context $\{A,B,C\}$.

The definition~(\ref{eq:MIE_proj_def}) admits an equivalent characterization in terms of projector commutators. As shown in SI, section B, the mutual information energy satisfies the exact correspondence
\begin{equation}\label{eq:E_comm_main}
1 - E(B;A,C) = \frac{1}{2d} \sum_{i,j} \|[\hat P^{A,B}_i, \hat P^{C,B}_j]\|_{\mathrm{HS}}^2,
\end{equation}
where $\|\cdot\|_{\mathrm{HS}}$ denotes the Hilbert--Schmidt norm. This identity reveals that the deviation of $E$ from unity is directly proportional to the total noncommutativity of the joint-eigenspace projectors. When all projectors commute, $E = 1$ and the context is noncontextual; increasing noncommutativity reduces $E$, signalling the presence of contextuality. The correspondence Eq.(\ref{eq:E_comm_main}) plays a central role in deriving the bounds presented in the following section.

\section{Operational contextuality measure}
\label{sec:OperationalMeasure}

While the mutual information energy $E(B;A,C)$ quantifies the intrinsic,
basis--independent geometric overlap within a single context, a complementary 
quantity is required to capture how this contextual behavior manifests 
\emph{operationally}—that is, at the level of measurement outcomes produced by 
a physical system in a given quantum state $\hat{\rho}$.  
To this end, we introduce a state--dependent, operational measure of 
contextuality defined over a family of contexts
\begin{equation}\label{eq:ContextSet}
  G=\{G_{\alpha}\}_{\alpha=1}^{N},\qquad
  G_{\alpha}=\{A_{\alpha},B_{\alpha},C_{\alpha}\},
\end{equation}
where each $G_{\alpha}$ represents a triad of observables whose associated operators satisfy $[\hat A_{\alpha},\hat B_{\alpha}]=0=[\hat B_{\alpha},\hat C_{\alpha}]$, defining a distinct measurement context.

\subsection{Definition}

In analogy with the role of commutators in the Robertson uncertainty relation\footnote{We use the Robertson form of the uncertainty relation, which generalizes the familiar Heisenberg bound.}~\cite{Robertson1929}, we quantify the operational signature of contextuality within each context
$G_{\alpha}$ by the magnitude of the commutator expectation value
\begin{equation}\label{eq:D_definition}
D(G_{\alpha},\hat\rho)
\;\coloneq\;
\bigl|\mathrm{Tr}\!\left([\hat A_{\alpha},\hat C_{\alpha}]\,\hat\rho\right)\bigr|.
\end{equation}
Summing over all contexts yields the global operational measure
\begin{equation}\label{eq:D_total_definition}
D(G,\hat\rho)
\;=\;
\sum_{\alpha=1}^{N}
D(G_{\alpha},\hat\rho).
\end{equation}
This quantity is state dependent and vanishes identically whenever all contexts consist of mutually commuting observables. Conversely, any nonzero contribution reflects an operationally accessible signature of contextuality for the state $\hat\rho$.

\subsection{Spectral decomposition bound}

A general upper bound for $D(G,\hat\rho)$ follows from the spectral decompositions of the observables. Let
\begin{equation}
\hat A_{\alpha}=\sum_{i} a_{\alpha i}\, \hat P_{\alpha i},
\qquad
\hat C_{\alpha}=\sum_{j} c_{\alpha j}\, \hat Q_{\alpha j},
\end{equation}
where $\hat P_{\alpha i} \equiv \hat P^{A_\alpha,B_\alpha}_{i}$ and $\hat Q_{\alpha j} \equiv \hat P^{C_\alpha,B_\alpha}_{j}$ are the projectors onto the joint eigenspaces of $(\hat A_\alpha, \hat B_\alpha)$ and $(\hat C_\alpha, \hat B_\alpha)$, respectively. By expanding the commutator in terms of these projectors and applying the Cauchy--Schwarz inequality in Hilbert--Schmidt space, one obtains the single-context bound
\begin{equation}\label{eq:D_single_bound_MainText}
D(G_{\alpha},\hat\rho)
\;\le\;
\kappa(A_{\alpha},C_{\alpha})\,
\bigl[1-E(B_{\alpha};A_{\alpha},C_{\alpha})\bigr]^{1/2},
\end{equation}
where the spectral prefactor is
\begin{equation}\label{eq:kappa_spectral_MainText}
\kappa(A_{\alpha},C_{\alpha})
\;\equiv\;
\sqrt{2d}\,
\biggl(\sum_{i} a_{\alpha i}^{2}\biggr)^{\!1/2}
\biggl(\sum_{j} c_{\alpha j}^{2}\biggr)^{\!1/2}.
\end{equation}
Summing over all contexts yields the global spectral bound
\begin{equation}\label{eq:Op_Global_Bound_MainText}
D(G,\hat\rho)
\;\le\;
\sum_{\alpha=1}^{N}
\kappa(A_{\alpha},C_{\alpha})\,
\bigl[1-E(B_{\alpha};A_{\alpha},C_{\alpha})\bigr]^{1/2}.
\end{equation}
The derivation of this bound, which relies on the projector--commutator correspondence Eq.(\ref{eq:E_comm_main}), is presented in SI, section C. Although completely general, it depends only on the eigenvalues of the observables and the mutual information energy, and therefore may overestimate the true operational contextuality for specific states.

\subsection{Purity-corrected bound}

The spectral bound~(\ref{eq:Op_Global_Bound_MainText}) can be tightened by retaining the purity of the quantum state. The purity is defined as
\begin{equation}\label{eq:purity_def}
\beta \;\equiv\; \|\hat\rho\|_{\mathrm{HS}}^2 = \mathrm{Tr}(\hat\rho^2),
\end{equation}
and satisfies $1/d \le \beta \le 1$, with $\beta = 1$ for pure states and $\beta = 1/d$ for the maximally mixed state. Incorporating this factor yields the purity-corrected bound
\begin{equation}\label{eq:purity_bound_MainText}
D(G_{\alpha},\hat\rho)
\;\le\;
\sqrt{\beta}\,\kappa(A_{\alpha},C_{\alpha})\,
\bigl[1-E(B_{\alpha};A_{\alpha},C_{\alpha})\bigr]^{1/2}.
\end{equation}
For mixed states with $\beta < 1$, this bound is strictly tighter than~(\ref{eq:D_single_bound_MainText}). The global purity-corrected bound is
\begin{equation}\label{eq:purity_global_MainText}
D(G,\hat\rho)
\;\le\;
\sqrt{\beta}\sum_{\alpha=1}^{N}
\kappa(A_{\alpha},C_{\alpha})\,
\bigl[1-E(B_{\alpha};A_{\alpha},C_{\alpha})\bigr]^{1/2}.
\end{equation}

\subsection{Operator norm bound}

An alternative state-independent bound follows from the duality between the operator norm and the trace norm. For any operator $\hat X$ and density matrix $\hat\rho$,
\begin{equation}
\bigl|\mathrm{Tr}(\hat X\,\hat\rho)\bigr| \;\le\; \|\hat X\|_{\mathrm{op}}\,\|\hat\rho\|_1 = \|\hat X\|_{\mathrm{op}},
\end{equation}
since $\|\hat\rho\|_1 = \mathrm{Tr}(\hat\rho) = 1$. Applied to the commutator, this yields
\begin{equation}\label{eq:op_norm_bound_MainText}
D(G_{\alpha},\hat\rho) \;\le\; \|[\hat A_{\alpha},\hat C_{\alpha}]\|_{\mathrm{op}}.
\end{equation}
Since $\|\hat X\|_{\mathrm{op}} \le \|\hat X\|_{\mathrm{HS}}$ for any operator, the operator norm bound is at least as tight as a direct application of the Cauchy--Schwarz inequality to $|\mathrm{Tr}([\hat A,\hat C]\,\hat\rho)|$. However, it does not always dominate the spectral bound~(\ref{eq:D_single_bound_MainText}), which follows a different derivation path through the projector structure and the MIE. The operator norm bound can be substantially tighter when the observables nearly commute despite having $E(B_\alpha; A_\alpha, C_\alpha) < 1$.

\subsection{Hybrid bound}

The purity-corrected spectral bound~(\ref{eq:purity_bound_MainText}) and the operator norm bound~(\ref{eq:op_norm_bound_MainText}) are complementary: the former captures the role of the MIE and state purity, while the latter is state-independent and often tight when the commutator is small. Taking the minimum yields the tightest estimate for a single context:
\begin{equation}\label{eq:hybrid_single_MainText}
D(G_{\alpha},\hat\rho) \;\le\; \min\Bigl\{ \|[\hat A_{\alpha},\hat C_{\alpha}]\|_{\mathrm{op}},\; \sqrt{\beta}\,\kappa(A_{\alpha},C_{\alpha})\,\bigl[1-E(B_{\alpha};A_{\alpha},C_{\alpha})\bigr]^{1/2} \Bigr\}.
\end{equation}
Summing over all $N$ contexts, the global hybrid bound reads
\begin{equation}\label{eq:hybrid_global_MainText}
D(G,\hat\rho) \;\le\; \sum_{\alpha=1}^{N} \min\Bigl\{ \|[\hat A_{\alpha},\hat C_{\alpha}]\|_{\mathrm{op}},\; \sqrt{\beta}\,\kappa_\alpha\,\bigl[1-E_\alpha\bigr]^{1/2} \Bigr\},
\end{equation}
where $\kappa_\alpha \equiv \kappa(A_{\alpha},C_{\alpha})$ and $E_\alpha \equiv E(B_{\alpha};A_{\alpha},C_{\alpha})$.

The global bound~(\ref{eq:hybrid_global_MainText}) is obtained by summing the single-context bounds and is mathematically valid for any collection of contexts. When contexts share observables—as in the KCBS scenario where each observable participates in two adjacent contexts—the bound remains correct but may be looser than if the contexts were independent. This is because the summation treats each context separately without accounting for correlations introduced by shared observables. The complete derivation of these bounds is given in SI, section C.

\subsection{Summary of bounds}

The hierarchy of bounds established above can be summarized as follows. The spectral bound,
\begin{equation}\label{eq:bound_spectral_summary}
D(G_\alpha,\hat\rho) \;\le\; \kappa_\alpha\,[1-E_\alpha]^{1/2},
\end{equation}
is the most general but loosest. The purity-corrected bound,
\begin{equation}\label{eq:bound_purity_summary}
D(G_\alpha,\hat\rho) \;\le\; \sqrt{\beta}\,\kappa_\alpha\,[1-E_\alpha]^{1/2},
\end{equation}
improves it for mixed states ($\beta < 1$). The operator norm bound,
\begin{equation}\label{eq:bound_op_summary}
D(G_\alpha,\hat\rho) \;\le\; \|[\hat A_\alpha,\hat C_\alpha]\|_{\mathrm{op}},
\end{equation}
provides an independent, state-independent estimate. Finally, the hybrid bound,
\begin{equation}\label{eq:bound_hybrid_summary}
D(G_\alpha,\hat\rho) \;\le\; \min\bigl\{\|[\hat A_\alpha,\hat C_\alpha]\|_{\mathrm{op}},\;\sqrt{\beta}\,\kappa_\alpha\,[1-E_\alpha]^{1/2}\bigr\},
\end{equation}
combines these to yield the tightest available constraint.

Taken together, these bounds clarify how the information--theoretic quantity $E(B;A,C)$ constrains the operational manifestations of contextuality. The geometric overlap encoded in the joint eigenspaces of $(A,B)$ and $(C,B)$ limits the possible size of the commutator expectation values, with the tightest constraint obtained by incorporating both the algebraic structure of each measurement context and the purity of the quantum state.

\section{Contextuality--dependent uncertainty relations}
\label{sec:App_Context_Uncert}

The bounds developed in Section~\ref{sec:OperationalMeasure} constrain the operational contextuality measure $D(G,\hat\rho)$ in terms of the mutual information energy. In this section we connect these bounds to the Robertson uncertainty relation, revealing a hierarchy that ties together the commutator structure, uncertainty products, and the geometric MIE quantity.

\subsection{Robertson lower bound}

For a quantum state $\hat\rho$, the variance of an observable $A_{\alpha}$ is 
\begin{equation}
(\Delta A_{\alpha})^2 = \mathrm{Tr}(\hat A_{\alpha}^2\,\hat\rho) - \bigl[\mathrm{Tr}(\hat A_{\alpha}\,\hat\rho)\bigr]^2.
\end{equation}
The Robertson relation~\cite{Robertson1929} provides a state--dependent lower bound on the product of uncertainties:
\begin{equation}\label{eq:Heisenberg_general}
\frac{1}{2}\bigl|\mathrm{Tr}([\hat A_{\alpha},\hat C_{\alpha}]\,\hat\rho)\bigr|
\;\le\;
(\Delta A_{\alpha})(\Delta C_{\alpha}),
\end{equation}
where the left-hand side quantifies the operational degree of incompatibility between $\hat{A}_\alpha$ and $\hat{C}_\alpha$ within the context $G_{\alpha}=\{A_{\alpha},B_{\alpha},C_{\alpha}\}$.

Recalling the definition~(\ref{eq:D_definition}) of the single-context operational measure, equation~(\ref{eq:Heisenberg_general}) can be rewritten as
\begin{equation}\label{eq:Robertson_D}
\frac{1}{2}\,D(G_{\alpha},\hat\rho) \;\le\; (\Delta A_{\alpha})(\Delta C_{\alpha}).
\end{equation}
Summing over all contexts in $G=\{G_{\alpha}\}_{\alpha=1}^{N}$ yields the collective lower bound
\begin{equation}\label{eq:Uncert_D}
\frac{1}{2}\,D(G,\hat\rho)
\;\le\; 
\sum_{\alpha=1}^N(\Delta A_{\alpha})(\Delta C_{\alpha}).
\end{equation}

\subsection{Hierarchy of bounds}

Combining the Robertson lower bound~(\ref{eq:Uncert_D}) with the information--theoretic upper bounds from Section~\ref{sec:OperationalMeasure}, we obtain a hierarchy relating the operational measure, the uncertainty products, and the mutual information energy.

From the spectral bound~(\ref{eq:Op_Global_Bound_MainText}), the chain of inequalities reads
\begin{equation}\label{eq:Hierarchy_spectral}
\frac{1}{2}\,D(G,\hat\rho)
\le
\sum_{\alpha=1}^N(\Delta A_{\alpha})(\Delta C_{\alpha}),
\quad
D(G,\hat\rho)
\le
\sum_{\alpha=1}^N\kappa_\alpha
\bigl[1-E_\alpha\bigr]^{1/2},
\end{equation}
where $\kappa_\alpha \equiv \kappa(A_{\alpha},C_{\alpha})$ and $E_\alpha \equiv E(B_{\alpha};A_{\alpha},C_{\alpha})$.

Incorporating the purity correction~(\ref{eq:purity_global_MainText}) tightens the upper bound for mixed states:
\begin{equation}\label{eq:Hierarchy_purity}
\frac{1}{2}\,D(G,\hat\rho)
\le
\sum_{\alpha=1}^N(\Delta A_{\alpha})(\Delta C_{\alpha}),
\quad
D(G,\hat\rho)
\le
\sqrt{\beta}\sum_{\alpha=1}^N\kappa_\alpha
\bigl[1-E_\alpha\bigr]^{1/2}.
\end{equation}

Finally, the hybrid bound~(\ref{eq:hybrid_global_MainText}) provides the tightest constraint:
\begin{equation}\label{eq:Hierarchy_hybrid}
\frac{1}{2}\,D(G,\hat\rho)
\le
\sum_{\alpha=1}^N(\Delta A_{\alpha})(\Delta C_{\alpha}),
\quad
D(G,\hat\rho)
\le
\sum_{\alpha=1}^N \min\bigl\{\|[\hat A_\alpha,\hat C_\alpha]\|_{\mathrm{op}},\;\sqrt{\beta}\,\kappa_\alpha\,[1-E_\alpha]^{1/2}\bigr\}.
\end{equation}

\subsection{Interpretation}

This hierarchy reveals the interplay between three distinct quantities:

\begin{itemize}
\item The \emph{Robertson lower bound} $\tfrac{1}{2}D(G,\hat\rho)$ sets the minimum uncertainty product consistent with the noncommutativity of the observables as experienced by the state $\hat\rho$.

\item The \emph{uncertainty products} $\sum_\alpha(\Delta A_\alpha)(\Delta C_\alpha)$ quantify the actual spread of measurement outcomes.

\item The \emph{information--theoretic upper bounds} constrain how large the operational contextuality can be, given the geometric structure encoded in the mutual information energy.
\end{itemize}

Although $E(B;A,C)$ does not directly bound the uncertainty products, it limits the operational size of the commutator expectations through $D(G,\hat\rho)$, thereby determining the contextuality--dependent scale on which the products $(\Delta A_{\alpha})(\Delta C_{\alpha})$ may vary.

\subsection{Transition to explicit examples}

At this point it is natural to ask whether these inequalities merely encode an abstract hierarchy, or whether they acquire a concrete operational meaning in a realistic quantum-mechanical system. To demonstrate this explicitly, we now turn to the simplest physical setting in which contextuality can arise: a single spin-1 particle. In this three-dimensional Hilbert space, the KCBS construction provides a canonical family of five dichotomic observables arranged along a pentagonal orthogonality graph. These observables not only exhibit the structural features identified above, but also allow the geometric quantity $E(B;A,C)$, the operational measure $D(G,\hat\rho)$, and the uncertainty products $(\Delta A)(\Delta C)$ to be computed in closed form and visualized geometrically via the Majorana--stellar representation. Thus, the KCBS scenario serves as an ideal testbed for the full framework developed in the preceding sections.

\section{Application to the KCBS Scenario}
\label{sec:KCBS}

Three-level quantum systems constitute the minimal Hilbert-space dimension in which contextuality can manifest~\cite{Gleason1957,Specker1960,Bell1966,KochenSpecker1967}. In this section we apply the framework developed above to the Klyachko--Can--Binicio\u{g}lu--Shumovsky (KCBS) scenario~\cite{KCBS}, where all quantities admit closed-form expressions and can be visualized geometrically through the Majorana--stellar representation.

\subsection{Spin-1 observables and pentagonal contexts}
\label{subsec:KCBS_observables}

For a spin-1 particle, the observable $\hat S_{\mathbf{k}} = \mathbf{\hat S}\cdot\mathbf{k}$ represents the spin component along a unit vector $\mathbf{k}$, with eigenvalues $m_s \in \{-1, 0, +1\}$. Following Klyachko et al.~\cite{KCBS}, one constructs the dichotomic observable
\begin{equation}\label{eq:Ak_def}
\hat A_{\mathbf{k}} = 2\hat S_{\mathbf{k}}^{2} - \hat I = \hat I - 2|0_{\mathbf{k}}\rangle\langle 0_{\mathbf{k}}|,
\end{equation}
which takes eigenvalues $\{+1, +1, -1\}$ with the $-1$ eigenspace spanned by the $m_s=0$ state $|0_{\mathbf{k}}\rangle$ along direction $\mathbf{k}$.

\begin{figure}[h!]
  \centering
  \begin{subfigure}[t]{0.45\textwidth}
      \includegraphics[width=\textwidth]{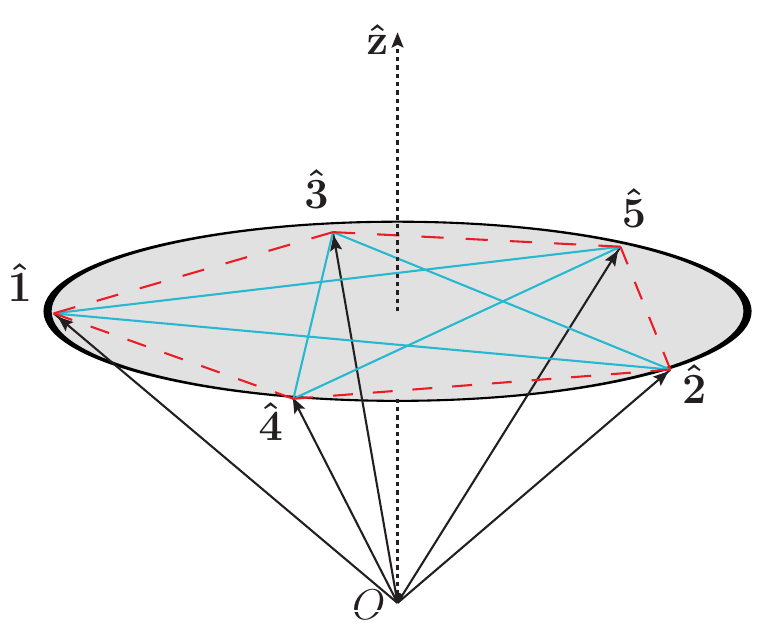}
 \caption{\normalsize{
KCBS configuration of five observables forming five overlapping contexts. The five directions $\mathbf{\hat 1}, \mathbf{\hat 2}, \mathbf{\hat 3}, \mathbf{\hat 4}, \mathbf{\hat 5}$ lie on a great circle with equal azimuthal spacing and satisfy the sequential orthogonality convention 
$\mathbf{\hat 1}\!\perp\!\mathbf{\hat 2}\!\perp\!\mathbf{\hat 3}\!\perp\!\mathbf{\hat 4}\!\perp\!\mathbf{\hat 5}\!\perp\!\mathbf{\hat 1}$.  
Each adjacent triplet $\{\hat{A}_{\mathbf{k}},\hat{A}_{\mathbf{k+1}},\hat{A}_{\mathbf{k+2}}\}$ defines a context in the KCBS set, and $\hat{A}_{\mathbf{k+5}}=\hat{A}_{\mathbf{k}}$.}}
 \label{fig:KCBS_10_z_copy}
 \end{subfigure}
~~
  \begin{subfigure}[t]{0.45\textwidth}
      \includegraphics[width=\textwidth]{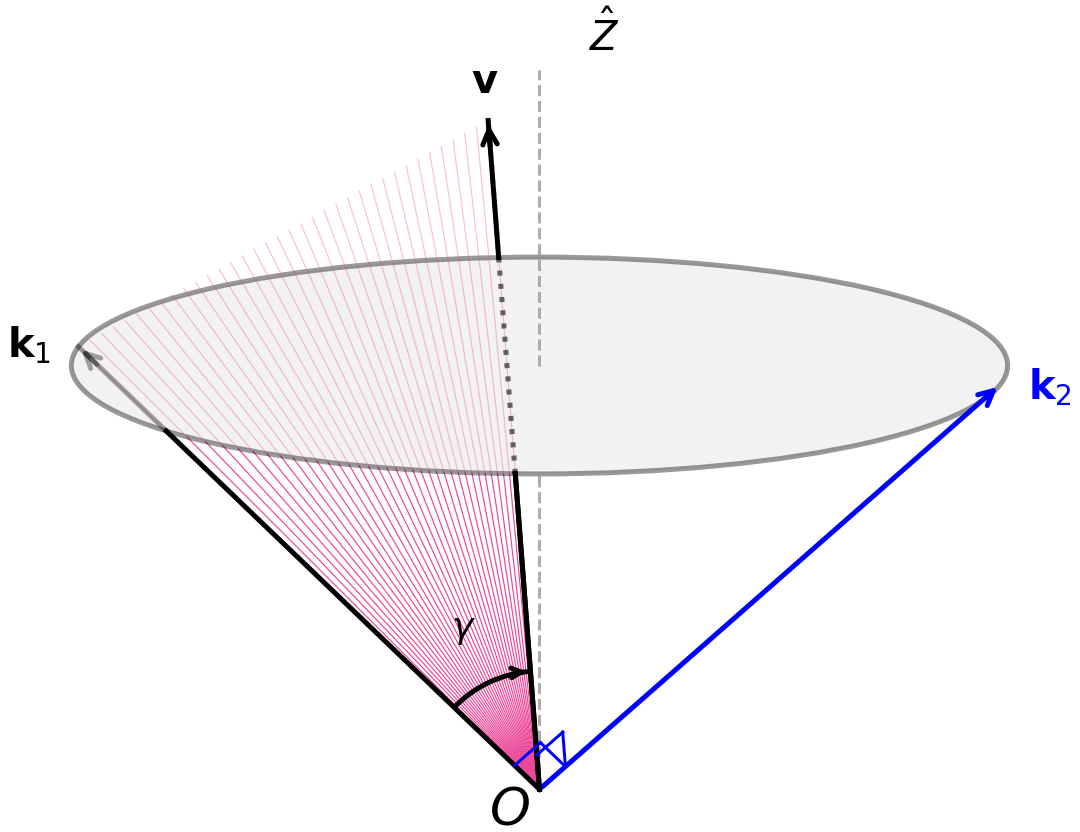}
 \caption{\normalsize{Orientation of a direction $\mathbf{k}_2$ perpendicular to the plane spanned by vectors $\mathbf{k}_1$ and $\mathbf{v}$. For $\gamma=\arccos\frac{\sqrt{5}-1}{2}$, direction $\mathbf{v}$ coincides with $\mathbf{\hat 3}-$axis shown on the left panel, and the set $\{\hat{A}_{\mathbf{1}},\hat{A}_{\mathbf{2}},\hat{A}_{\mathbf{3}}\}$ becomes one of the five KCBS contexts.}}
      \label{fig:kcbs_modified}
  \end{subfigure}
\caption{\normalsize{KCBS configuration and local orthogonal triad.}}
  \label{fig:KCBS}
\end{figure}

The KCBS construction employs five directions $\mathbf{\hat 1}, \ldots, \mathbf{\hat 5}$ 
arranged symmetrically about the $z$-axis, satisfying the cyclic orthogonality condition
\begin{equation}\label{eq:KCBS_orthogonality}
\mathbf{k} \perp \mathbf{k+1}, \quad \pmod 5.
\end{equation}
Geometrically, these five unit vectors lie on a cone of fixed polar angle
\[
\theta_{\mathrm{KCBS}}
=
\arcsin\!\left(\frac{1}{\sqrt{2}\cos(\pi/10)}\right)
\approx 63.44^\circ,
\]
with azimuthal angles
\[
\varphi_\alpha = (\alpha-1)\frac{6\pi}{5}, 
\qquad \alpha\pmod5,
\]
as shown in Fig.~\ref{fig:KCBS_10_z_copy}.  
Together, this parametrization realizes the cyclic orthogonality in~(\ref{eq:KCBS_orthogonality}) and fixes the geometry of the KCBS pentagon.  
The angle $\gamma$ between non-adjacent directions $\mathbf{k}$ and $\mathbf{k+2}$ is then determined purely by this pentagonal symmetry to be
\begin{equation}\label{eq:gamma_KCBS}
\cos\gamma = \frac{\sqrt{5}-1}{2} \approx 0.618,
\qquad
\gamma \approx 51.83^{\circ}.
\end{equation}

This configuration defines five overlapping contexts
\begin{equation}\label{eq:KCBS_contexts}
G_\alpha = \{A_{\alpha-1}, A_\alpha, A_{\alpha+1}\}, \qquad \alpha\pmod5,
\end{equation}
where within each context the central observable $A_\alpha$ commutes with both neighbors while the outer pair $\{A_{\alpha-1}, A_{\alpha+1}\}$ do not commute (see Fig.~\ref{fig:KCBS_10_z_copy}).

\subsection{Explicit evaluation of the contextuality measures}
\label{subsec:KCBS_evaluation}
Before turning to the KCBS scenario itself, we first present an explicit non-KCBS example to illustrate how the projection-based mutual information energy operates in both degenerate and nondegenerate settings.
This example, involving $\{S_{\mathbf{k}} \}$ and $\{A_{\mathbf k}\}$, is not part of the KCBS construction; its purpose is solely to clarify the behaviour of the MIE under changes of measurement structure and spectral degeneracy.
Once this illustrative case is established, we then return to the KCBS pentagon and evaluate all quantities of interest— E(B;A,C), the spectral prefactors, the operator-norm bounds, and the total operational contextuality $D(G,\hat{\rho})$—for the five contextual triples $G_\alpha$.

For a direction $\mathbf{v}$ lying on the plane spanned by $\mathbf{k}_1$ and $\mathbf{k}_3$, let $\gamma$ be the angle between $\mathbf{k}_1$ and $\mathbf{v}$ (as in Fig.~\ref{fig:kcbs_modified}).  
The mutual information energies corresponding to the contexts $\{S_{\mathbf{k}_1},S^2,S_{\mathbf{v}}\}$ and $\{A_{\mathbf{k}_1},A_{\mathbf{k}_2},A_{\mathbf{v}}\}$ are found to be
\begin{equation}\label{eq:MIE_closed_form_S}
E(S^2;S_{\mathbf{k}_1},S_{\mathbf{v}})=\frac{5-2\cos^2\gamma+9\cos^4\gamma}{12},
\end{equation}
and
\begin{equation}\label{eq:MIE_closed_form_A}
E\big(A_{\mathbf{k}_2};A_{\mathbf{k}_1},A_{\mathbf{v}}\big)
=\frac{3-4\cos^2\gamma+4\cos^4\gamma}{3}.
\end{equation}
Although the operators 
$\hat{A}_{\mathbf{k}_1}\hat{A}_{\mathbf{k}_2}$ 
and 
$\hat{A}_{\mathbf{v}}\hat{A}_{\mathbf{k}_2}$ 
can share the same eigenbases as 
$\hat S_{\mathbf{k}_1}$ 
and 
$\hat S_{\mathbf{v}}$, 
respectively, degeneracies in the measurements of $\hat{A}_{\mathbf{k}}$ yield distinct contextuality values for the two sets 
$\{\hat{A}_{\mathbf{k}_1},\hat{A}_{\mathbf{k}_2},\hat{A}_{\mathbf{v}}\}$ 
and 
$\{S_{\mathbf{k}_1},S^2,S_{\mathbf{v}}\}$.  
All MIE values are computed using the joint projectors of the commuting pairs onto the eigenspaces, as discussed in  SI, section B, thereby avoiding artefacts due to degeneracy or basis choice.

For each KCBS context $G_\alpha$, $\gamma$ is given in Eq.(\ref{eq:gamma_KCBS}) and the mutual information energy~(\ref{eq:MIE_proj_def}) evaluates to
\begin{equation}\label{eq:MIE_KCBS}
E(A_\alpha; A_{\alpha-1}, A_{\alpha+1}) = \frac{3 - 4\cos^2\gamma + 4\cos^4\gamma}{3} \approx 0.685.
\end{equation}
The deviation from unity, $1 - E \approx 0.315$, quantifies the intrinsic contextuality of each KCBS context. This value is uniform across all five contexts by the pentagonal symmetry. Note that $E = 1$ when $\cos\gamma = 0$ or $\pm 1$, corresponding to orthogonal or parallel directions; in these limits the outer observables in the set~(\ref{eq:KCBS_contexts}) commute and the context becomes noncontextual, as expected.

The spectral prefactor~(\ref{eq:kappa_spectral_MainText}) for dichotomic observables with eigenvalues $\{+1, +1, -1\}$ is
\begin{equation}\label{eq:kappa_KCBS}
\kappa(A_{\alpha-1}, A_{\alpha+1}) = \sqrt{2d}\,\biggl(\sum_i a_i^2\biggr)^{\!1/2}\biggl(\sum_j c_j^2\biggr)^{\!1/2} = \sqrt{6} \times 3 = 3\sqrt{6} \approx 7.35,
\end{equation}
where $d = 3$ and $\sum_i a_i^2 = \sum_j c_j^2 = 1 + 1 + 1 = 3$.\\
Combining these, the spectral bound~(\ref{eq:D_single_bound_MainText}) for a single context gives
\begin{equation}\label{eq:spectral_KCBS}
D(G_\alpha, \hat\rho) \le \kappa\sqrt{1-E} \approx 7.35 \times 0.561 \approx 4.12.
\end{equation}

The commutator of the outer observables in the context $G_\alpha$~(\ref{eq:KCBS_contexts}) can be computed directly. Since $[\hat A_{\mathbf{k}}, \hat A_{\mathbf{k}'}] = 4[\hat P_{0_{\mathbf{k}}}, \hat P_{0_{\mathbf{k}'}}]$ where $\hat P_{0_{\mathbf{k}}} = |0_{\mathbf{k}}\rangle\langle 0_{\mathbf{k}}|$, and the projector commutator has operator norm $\|[\hat P, \hat Q]\|_{\mathrm{op}} = \cos\gamma\sin\gamma$ for rank-1 projectors with overlap $|\langle 0_{\mathbf{k}}|0_{\mathbf{k}'}\rangle|^2 = \cos^2\gamma$, we obtain
\begin{equation}\label{eq:op_norm_KCBS}
\|[\hat A_{\alpha-1}, \hat A_{\alpha+1}]\|_{\mathrm{op}} = 4\sqrt{\sqrt{5}-2} \approx 1.94.
\end{equation}
This operator norm bound is substantially tighter than the spectral bound, differing by a factor exceeding two.

For pure states ($\beta = 1$), the hybrid bound~(\ref{eq:hybrid_single_MainText}) therefore reduces to the operator norm bound for all KCBS contexts. Summing over all five contexts yields the global bound
\begin{equation}\label{eq:global_bound_KCBS}
D(G, \hat\rho) \le 5 \times 1.94 \approx 9.72.
\end{equation}

\subsection{Geometric visualization via the Majorana--stellar representation}
\label{subsec:MSR}

The Majorana--stellar representation~\cite{MajoranaSR1932} provides a geometric framework for visualizing spin-$s$ states as constellations on the Bloch--Poincar\'e sphere. Beyond visualization, this representation offers significant computational advantages: inner products, expectation values, and projector overlaps reduce to elementary functions of angles between stellar directions, bypassing explicit matrix algebra. This geometric formalism has also been used in detailed analyses of spin-1 state geometry, notably in the work of Aravind~\cite{Aravind2017}.
The Majorana--stellar representation thus furnishes a common geometric platform on which both the operational contextuality measure $D(G,\hat\rho)$ and the uncertainty products $(\Delta A)(\Delta C)$ can be analyzed and visualized. For spin-1, the three basis states $|{+}1_{\mathbf{k}}\rangle$, $|0_{\mathbf{k}}\rangle$, $|{-}1_{\mathbf{k}}\rangle$ are each represented by a pair of Majorana stars, and any pure state is a superposition of these basis vectors. The eigenstates of $\hat S_{\mathbf{k}}$ take the forms
\begin{equation}\label{eq:MSR_eigenstates}
|{+}1_{\mathbf{k}}\rangle \equiv |{+k}{+k}\rangle, \quad
|0_{\mathbf{k}}\rangle \equiv \tfrac{1}{\sqrt{2}}\bigl(|{+k}{-k}\rangle + |{-k}{+k}\rangle\bigr), \quad
|{-}1_{\mathbf{k}}\rangle \equiv |{-k}{-k}\rangle,
\end{equation}
where the notation $|{\pm k}\rangle$ denotes spin-$\tfrac{1}{2}$ states along direction $\pm\mathbf{k}$. The $m_s = 0$ state thus corresponds to two antipodal stars aligned with $\mathbf{k}$.

A general pure spin-1 state with Majorana directions $\mathbf{m}$ and $\mathbf{n}$ can be written as
\begin{equation}\label{eq:general_spin1}
|\chi\rangle = \frac{1}{\sqrt{3 + \mathbf{m}\cdot\mathbf{n}}}\bigl(|{+m}{+n}\rangle + |{+n}{+m}\rangle\bigr),
\end{equation}
and its overlap with the $m_s=0$ state along $\mathbf{k}$ determines the expectation value $\langle\chi|\hat P_{0_{\mathbf{k}}}|\chi\rangle$.

\begin{figure}[h!]
  \centering
  \begin{subfigure}[t]{0.31\textwidth}
      \includegraphics[width=\textwidth]{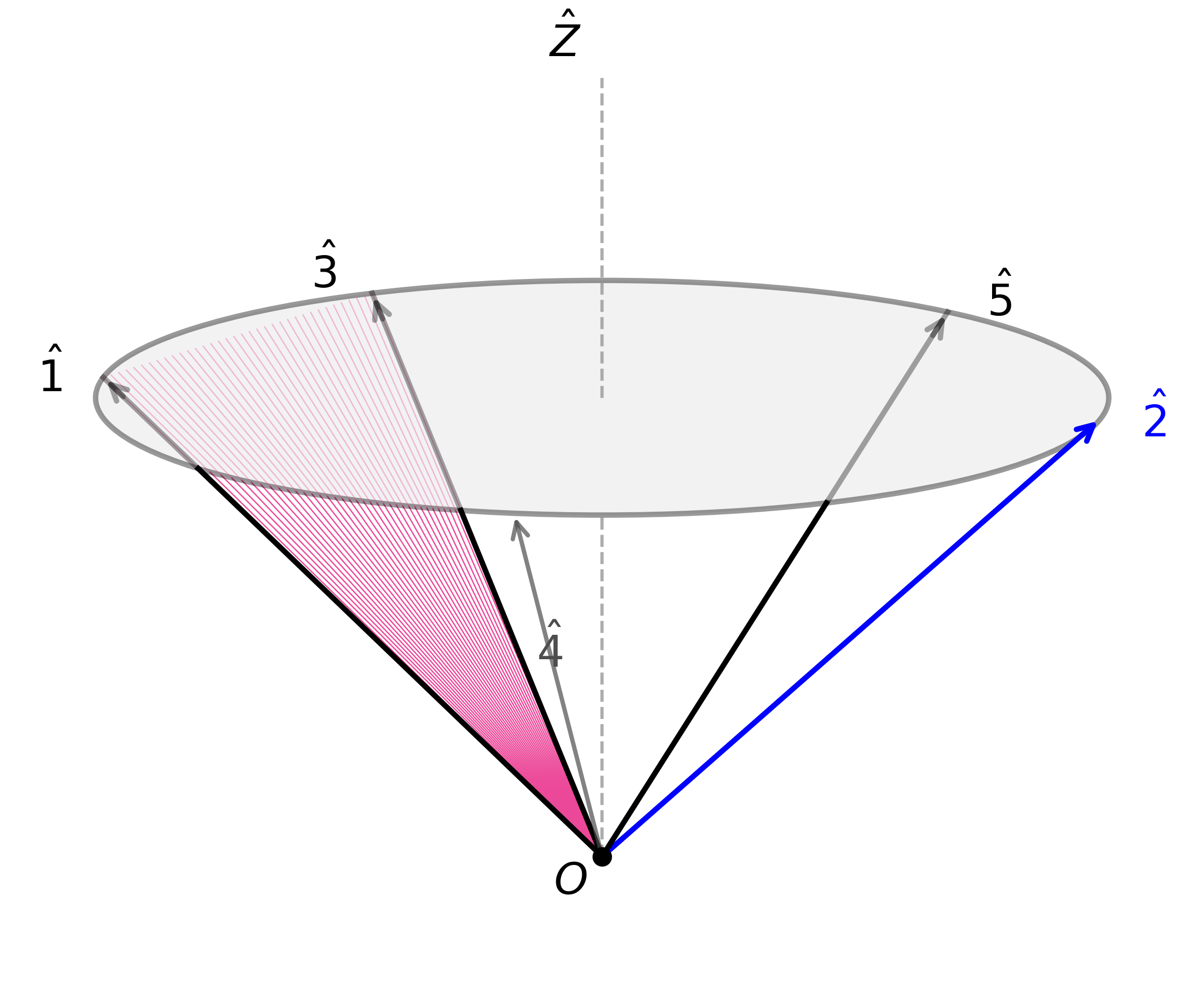}
      \caption{\normalsize{The pink surface, perpendicular to the axis $\mathbf{\hat 2}$, is spanned by axes $\mathbf{\hat 1}$ 
      and $\mathbf{\hat 3}$.}}
      \label{fig:KCBS_Surface01}
  \end{subfigure}
  ~
  \begin{subfigure}[t]{0.31\textwidth}
      \includegraphics[width=\textwidth]{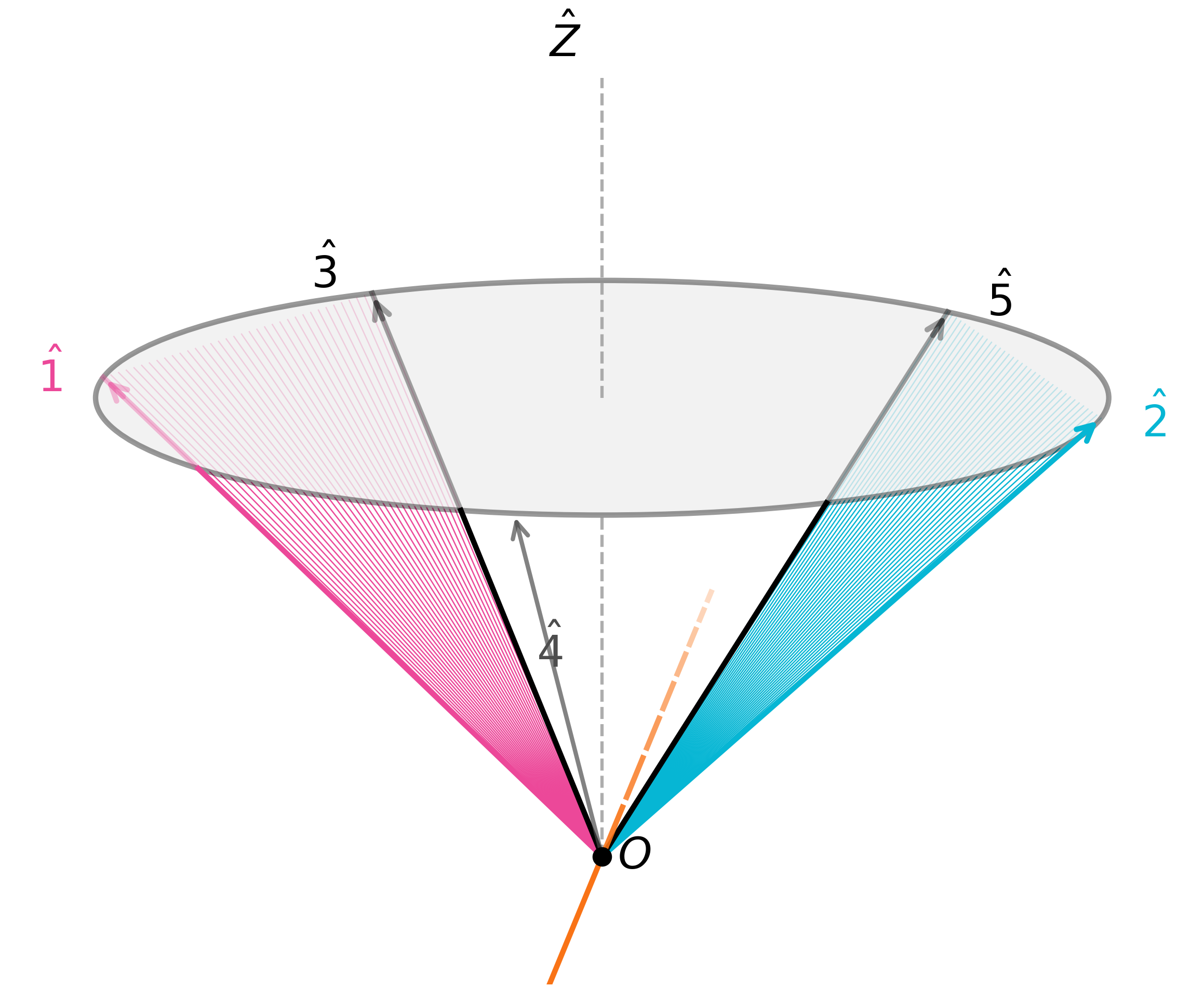}
      \caption{\normalsize{States represented on the intersection line of the pink and blue surfaces, shown here by \textcolor{orange}{orange} line, exhibit 
      maximum contextuality for both $A_{\mathbf 1}$ and 
      $A_{\mathbf 2}$.}}
      \label{fig:KCBS_Surfaces}
  \end{subfigure}
  ~
  \begin{subfigure}[t]{0.31\textwidth}
      \includegraphics[width=\textwidth]{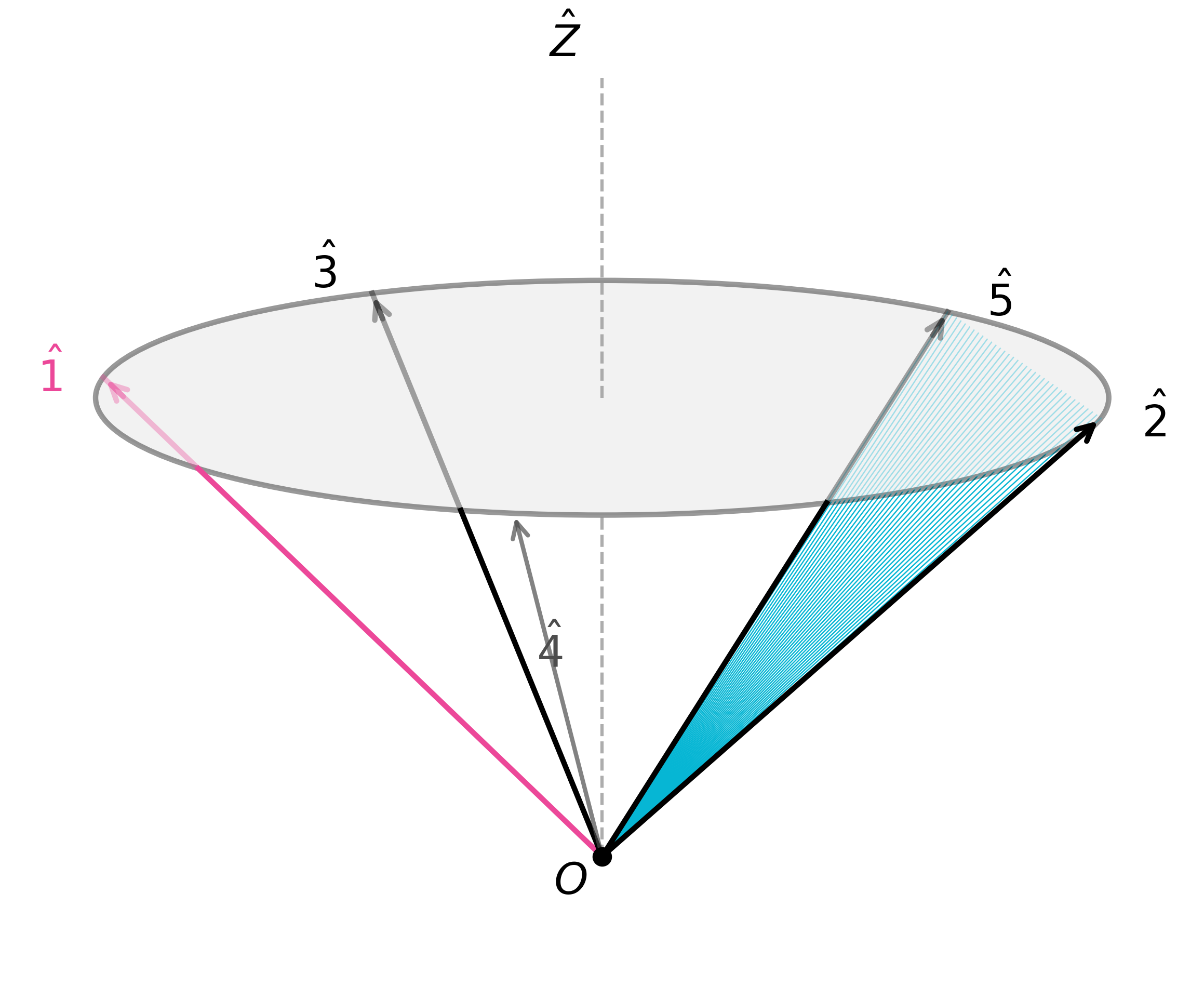}
      \caption{\normalsize{The blue surface, perpendicular to the axis $\mathbf{\hat 1}$, is spanned by axes $\mathbf{\hat 2}$ and $\mathbf{\hat 5}$.}}
      \label{fig:KCBS_Surface02}
  \end{subfigure}
  \caption{\normalsize{Maximum contextuality for an observable $A_{\mathbf k}$ of a spin-1 system occurs for the states that can be represented on the surface perpendicular to direction $\mathbf k$ in the Majorana–stellar representation. This surface is the one spanned by the directions along which the other two observables of the same context are defined. Two such surfaces and their intersection are shown. All the 5 directions $\mathbf{\hat 1}-\mathbf{\hat 5}$ correspond to those in the KCBS pentagram of figure~\ref{fig:KCBS_10_z_copy}.}}
  \label{fig:KCBS_Surfaces_3}
\end{figure}

The uncertainty of the dichotomic observable $A_{\mathbf{k}}$ in state $|\chi\rangle$ is
\begin{equation}\label{eq:variance_Ak}
(\Delta A_{\mathbf{k}})^2 = 4\,p_{\mathbf{k}}(1 - p_{\mathbf{k}}), \qquad p_{\mathbf{k}} = \langle\chi|\hat P_{0_{\mathbf{k}}}|\chi\rangle,
\end{equation}
and is maximized when $p_{\mathbf{k}} = \tfrac{1}{2}$. For spin-1, the Majorana--stellar representation yields a three-dimensional Euclidean-like visualization of the Hilbert space. Within this geometric picture, the maximum-uncertainty condition defines a plane perpendicular to $\mathbf{k}$, spanned by the directions of the other two observables in the same context (see Fig.~\ref{fig:KCBS_Surfaces_3}). States that exhibit maximum uncertainty for $A_{\mathbf{k}}$ has
Majorana constellation lying on this plane .

\begin{figure}[h]
    \centering
    \includegraphics[width=0.5\linewidth]{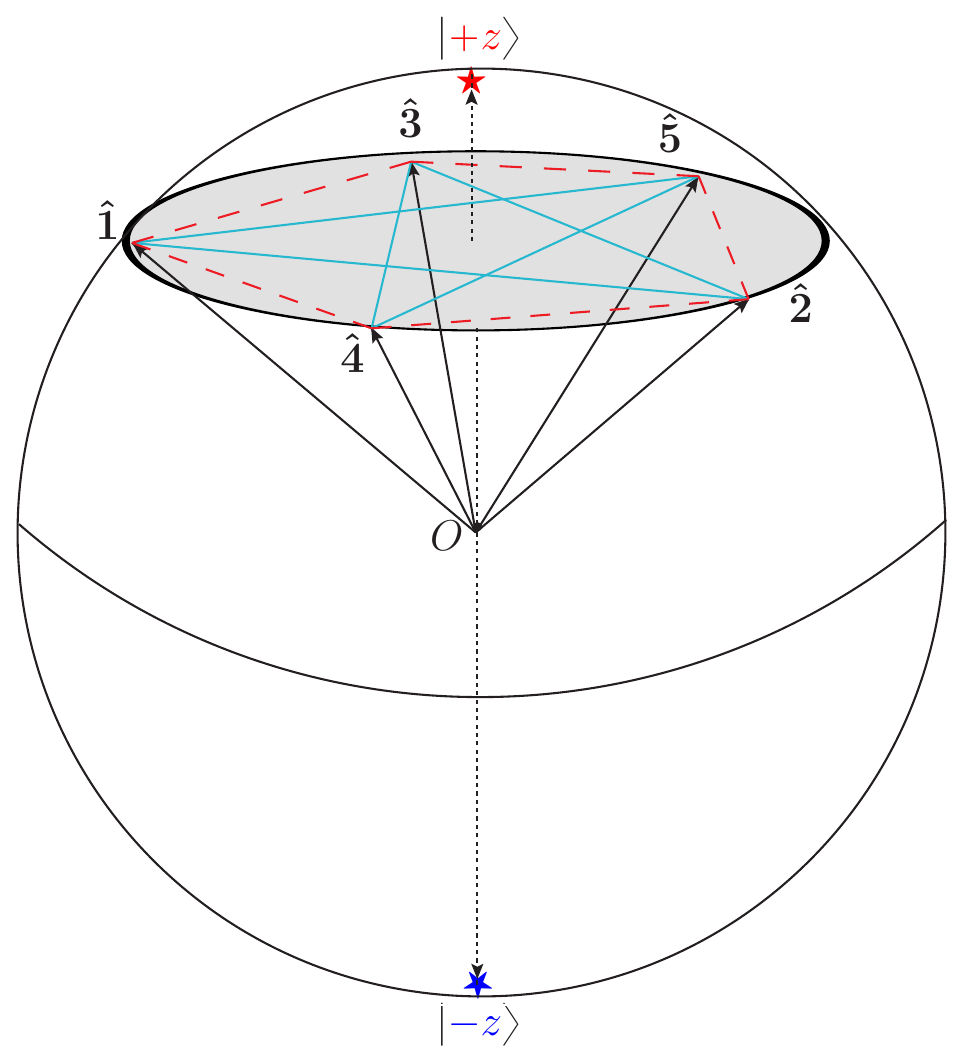}
    \caption{\normalsize{KCBS pentagram and the state $|0_{\mathbf{z}}\rangle$ represented by a Majorana constellation composed of two stars, located on the BP sphere; $|0_{\mathbf{z}}\rangle=\frac{1}{\sqrt 2}(|{\color{red}+z}{\color{blue}-z}\rangle+ |{\color{blue}-z}{\color{red}+z}\rangle)
\equiv|\hat{Z}\rangle$.}}
    \label{fig:KCBS_10_O}
\end{figure}

For two different directions $\mathbf{k}$ and $\mathbf{k'}$, the associated maximum-uncertainty planes intersect along a curve (Fig.~\ref{fig:KCBS_Surfaces}). However, for three or more contexts, the corresponding planes do not generically share a common line; hence true maximum uncertainty cannot be achieved simultaneously for all contexts. Instead, one seeks the state that optimizes the sum of uncertainty products. For the full KCBS family of five contexts, this optimization yields a unique state: $|0_{\mathbf{z}}\rangle$, the $m_s = 0$ eigenstate along the symmetry axis (Fig.~\ref{fig:KCBS_10_O}; see also Table~\ref{tab:SmaxHierarchy2}).

Further geometric characterizations of spin-1 states in terms of orthonormal triads, including explicit coefficient formulas for the decomposition $|\chi\rangle = K|\hat K\rangle + K_1|\hat K_1\rangle + K_2|\hat K_2\rangle$, are developed in SI, section D.

\subsection{States achieving extremal uncertainty products}
\label{subsec:extremal_states}

The state $|0_{\mathbf{z}}\rangle$ achieves the optimal sum of uncertainty products
\begin{equation}\label{eq:max_uncertainty_product}
\sum_{\alpha=1}^{5} (\Delta A_{\alpha-1})(\Delta A_{\alpha+1}) = 4(\sqrt{5} - 1) \approx 4.94,
\end{equation}
with each context contributing equally: $(\Delta A_{\alpha-1})(\Delta A_{\alpha+1}) = \tfrac{4(\sqrt{5}-1)}{5} \approx 0.99$ per context.

A remarkable feature of this maximizing state---and more generally of all states of the form $|0_{\hat{\mathbf{u}}}\rangle$ for any axis $\hat{\mathbf{u}}$---is that the operational contextuality measure vanishes:
\begin{equation}\label{eq:D_vanishes}
D(G, |0_{\hat{\mathbf{u}}}\rangle\langle0_{\hat{\mathbf{u}}}|) = \sum_{\alpha=1}^{5} \bigl|\langle 0_{\hat{\mathbf{u}}}|[\hat A_{\alpha-1}, \hat A_{\alpha+1}]|0_{\hat{\mathbf{u}}}\rangle\bigr| = 0.
\end{equation}
This follows from the quadrupolar structure of the KCBS observables: since $\hat A_{\mathbf{k}} = 2(\mathbf{\hat S}\cdot\mathbf{k})^2 - \hat I$ transforms covariantly under rotations, and states $|0_{\hat{\mathbf{u}}}\rangle$ possess rotational symmetry about $\hat{\mathbf{u}}$ with $\langle\mathbf{\hat S}\rangle = 0$, the antisymmetric part of any operator product averages to zero.

Consequently, the Robertson lower bound~(\ref{eq:Robertson_D}) becomes trivial for these states:
\begin{equation}\label{eq:Robertson_trivial}
\frac{1}{2}D(G, |0_{\hat{\mathbf{u}}}\rangle\langle0_{\hat{\mathbf{u}}}|) = 0 \le \sum_{\alpha}(\Delta A_{\alpha-1})(\Delta A_{\alpha+1}).
\end{equation}
The bound is satisfied but provides no constraint on the uncertainty products. This illustrates that maximum uncertainty and maximum operational contextuality are achieved by \emph{different} quantum states in the KCBS scenario.

In contrast, generic states such as $|{\pm}1_{\mathbf{z}}\rangle$ exhibit nonzero operational contextuality. For these eigenstates of $\hat S_z$, numerical evaluation gives $D(G, |{\pm}1_{\mathbf{z}}\rangle\langle {\pm}1_{\mathbf{z}}|) \approx 6.50$, which represents approximately $67\%$ of the global operator norm bound $5 \times 1.94 \approx 9.72$. The Robertson lower bound then yields $\tfrac{1}{2}D \approx 3.25$, while the actual uncertainty product is $\sum_\alpha(\Delta A_{\alpha-1})(\Delta A_{\alpha+1}) = 4.00$, leaving a gap of approximately $0.75$.

\subsection{Summary of the KCBS analysis}
\label{subsec:KCBS_summary}

The KCBS scenario provides a concrete illustration of the full framework. The mutual information energy $E \approx 0.685$ quantifies the intrinsic contextuality of each pentagonal context, while the operator norm bound $\|[\hat A, \hat C]\|_{\mathrm{op}} \approx 1.94$ provides the tightest constraint on the operational measure $D(G_\alpha, \hat\rho)$---substantially sharper than the spectral bound of $4.12$ per context.

The Majorana--stellar representation reveals that the state achieving the optimal sum of uncertainty products is unique: $|0_{\mathbf{z}}\rangle$, which attains $\sum_\alpha(\Delta A_{\alpha-1})(\Delta A_{\alpha+1}) = 4(\sqrt{5}-1) \approx 4.94$ yet has vanishing operational contextuality $D = 0$. This decoupling between uncertainty products and commutator expectations demonstrates that the Robertson inequality, while always valid, can become uninformative precisely at the states of greatest physical interest.

The hierarchy of bounds developed in Sections~\ref{sec:OperationalMeasure}--\ref{sec:App_Context_Uncert} is thus fully verified in the KCBS setting, with the operator norm bound emerging as the relevant constraint for this geometry.

\section{Conclusion}
\label{sec:Conclusion}

Contextuality lies at the heart of quantum theory~\cite{Heisenberg1925,Nobel1932,SakuraiNapolitano2017} and constitutes a key resource for quantum information processing and computation~\cite{HeywoodRedhead1983, PasquinucciPeres2000, Spekkens2008, AharonVaidman2008, Svozil2009, Cabello2010, WAEGELL2013546, abramsky_et_al:LIPIcs:2015:5416, Raussendorf2013, Frembs2018}. It unifies several of the most striking features of quantum mechanics, including the incompatibility of measurements~\cite{Bell1966, PopescuRohrlich1994}, entanglement~\cite{Schrodinger1935, EPR1935, Bohm1951}, and nonlocality~\cite{Bell1964, Stapp1975, BellAspect2004, GISIN1996151}. A complete understanding and systematic quantification of contextuality is therefore essential, both for foundational reasons and for practical applications in quantum technologies.

In this work we have proposed an information--theoretic framework for quantifying Kochen--Specker contextuality. Two complementary measures were introduced: the mutual information energy $E(B;A,C)$, a state--independent quantity that captures the geometric overlap between the joint eigenspaces of $(A,B)$ and $(C,B)$ within each context, quantifying how the incompatible observables $A$ and $C$ relate to one another; and the operational measure $D(G,\hat\rho)$, a state--dependent quantity that reflects the contextual behavior of observables through commutator expectation values. The mutual information energy, inspired by Onicescu's information energy, equals unity for noncontextual configurations and decreases as contextuality increases. The operational measure provides an experimentally accessible signature, directly tied to the Robertson uncertainty relation.

We established a hierarchy of bounds connecting these measures to the uncertainty products of incompatible observables. The spectral bound relates $D$ to the mutual information energy through the prefactor $\kappa = \sqrt{2d}\,(\sum_i a_i^2)^{1/2}(\sum_j c_j^2)^{1/2}$, while a purity correction tightens this bound for mixed states. More significantly, the operator norm bound $D \le \|[\hat A, \hat C]\|_{\mathrm{op}}$ can provide a tighter constraint than the spectral estimate, as demonstrated explicitly for the KCBS scenario. The hybrid bound, taking the minimum of spectral and operator norm estimates, provides the most refined upper limit available from these methods.

Application to the KCBS scenario---the minimal contextuality configuration in a three--dimensional Hilbert space---yielded explicit closed--form expressions for all quantities. The pentagonal geometry fixes the mutual information energy at $E \approx 0.685$ for each context, with the operator norm bound ($\approx 1.94$ per context) proving more than twice as tight as the spectral bound ($\approx 4.12$ per context). The Majorana--stellar representation provided both computational advantages and geometric insight: for spin-1, it yields a three-dimensional Euclidean-like visualization in which maximum--uncertainty states for a given observable lie on the plane perpendicular to that observable's direction. Since three or more maximum-uncertainty planes cannot share a common intersection, simultaneous maximum uncertainty is unattainable; the state $|0_{\mathbf{z}}\rangle$ instead optimizes the sum of uncertainty products, achieving the global maximum $\sum_\alpha (\Delta A_{\alpha-1})(\Delta A_{\alpha+1}) = 4(\sqrt{5}-1)$.

A notable finding concerns the relationship between operational contextuality and uncertainty. States of the form $|0_{\hat{\mathbf{u}}}\rangle$---which achieve the optimal sum of uncertainty products---exhibit vanishing operational contextuality ($D = 0$) due to the quadrupolar symmetry of the KCBS observables. For these states the Robertson inequality becomes trivial, providing no constraint on the uncertainties. In contrast, generic states such as $|{\pm}1_{\mathbf{z}}\rangle$ display substantial operational contextuality ($D \approx 6.50$) with a nontrivial Robertson lower bound. This demonstrates that maximum uncertainty and maximum operational contextuality are achieved by distinct quantum states, revealing a subtle interplay between these fundamental aspects of quantum mechanics. Whether this decoupling persists in other contextuality scenarios---where the quadrupolar observable structure and rotational symmetry specific to KCBS may not hold---warrants systematic investigation.

The geometric approach developed here, combining the mutual information energy with the Majorana--stellar visualization, offers a unified perspective on contextuality in finite--dimensional quantum systems. Extensions to higher--spin systems and alternative contextuality scenarios represent promising directions for future investigation.

\section*{Acknowledgments}
Part of this work was carried out at Bilimler Köyü, Foça, Türkiye.

\begin{figure}[h!]
  \centering
  \begin{subfigure}[h!]{0.3\textwidth}
      \includegraphics[width=\textwidth]{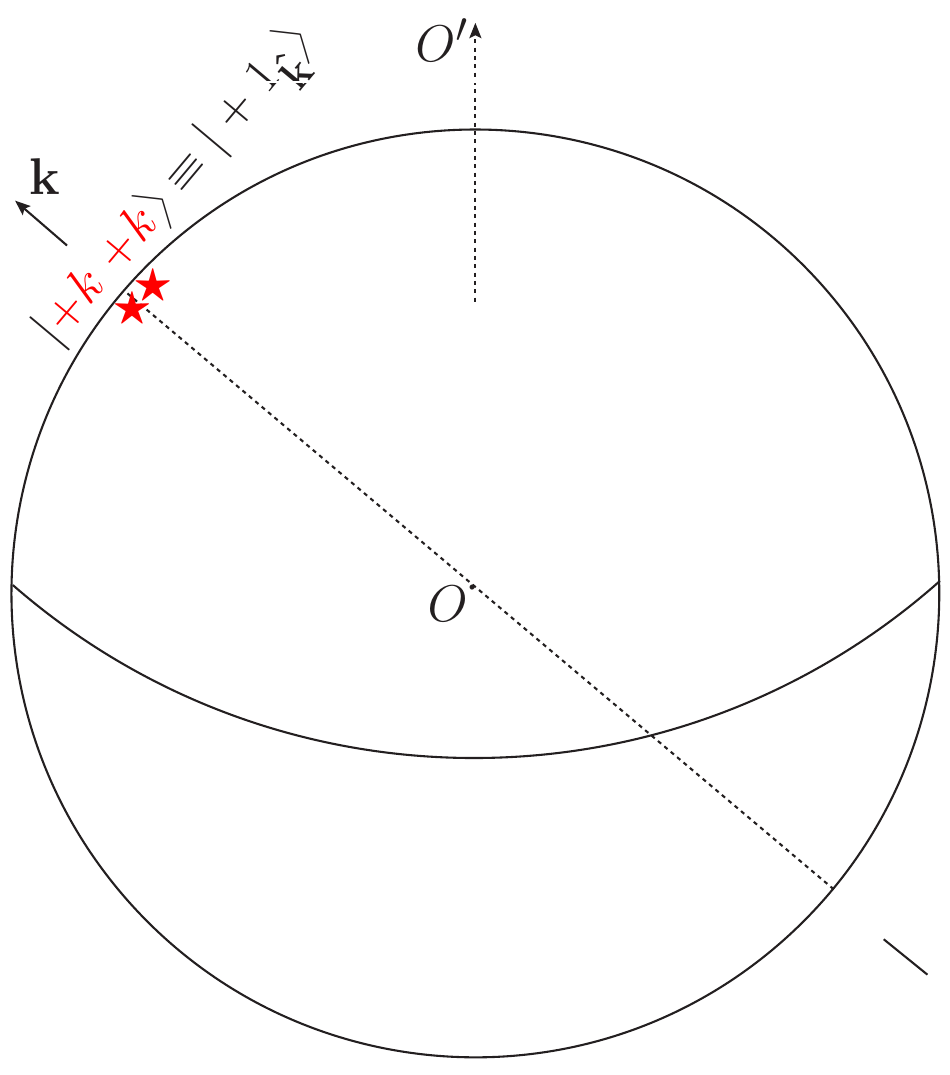}
 \caption{$|+1_{\mathbf{k}}\rangle\equiv|{\color{red}{+k+k}}\rangle$.}
      \label{fig:MSR1}
  \end{subfigure}
~
  \begin{subfigure}[h!]{0.3\textwidth}
      \includegraphics[width=\textwidth]{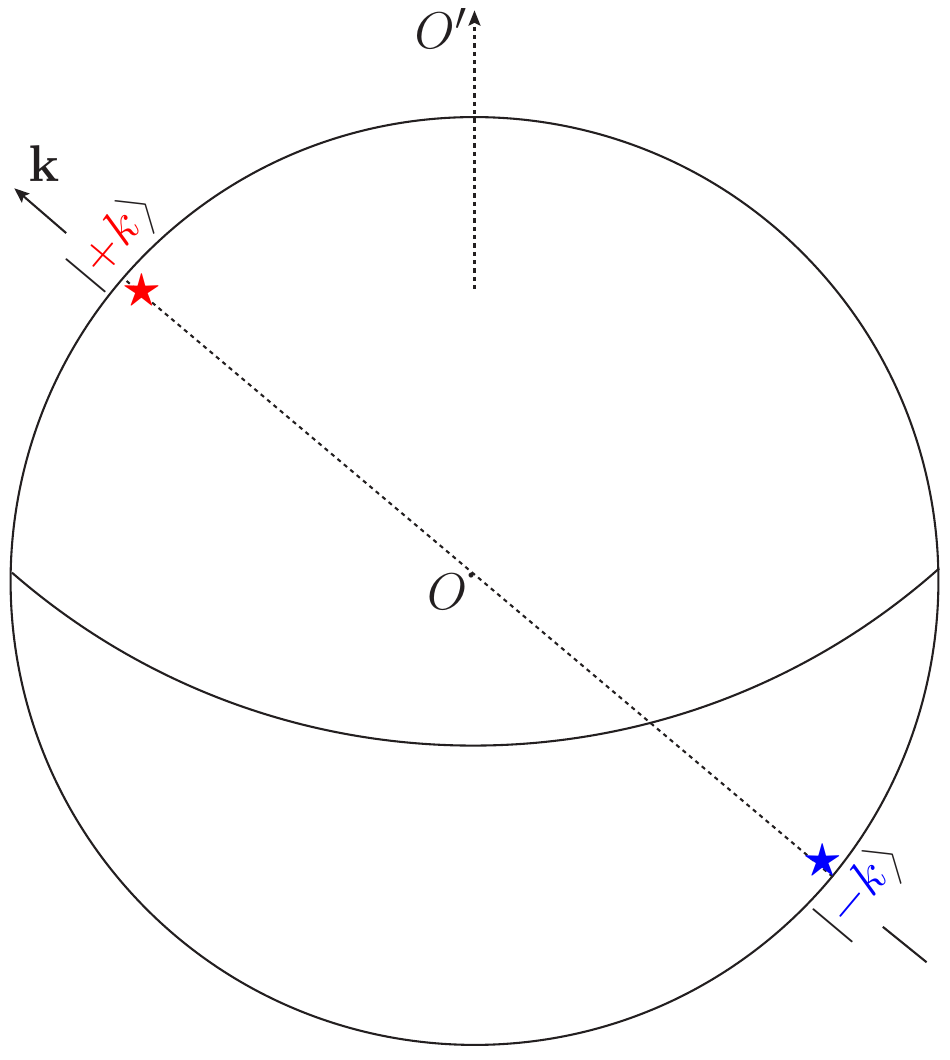}
 \caption{$|0_{\mathbf{k}}\rangle\equiv\frac{1}{\sqrt2}(|{\color{red}+k}{\color{blue}-k}\rangle+\\ \qquad\:\:\:\quad\qquad|{\color{blue}-k}{\color{red}+k}\rangle).$}
      \label{fig:MSR2}
  \end{subfigure}
~
  \begin{subfigure}[h!]{0.3\textwidth}
      \includegraphics[width=1.12\textwidth]{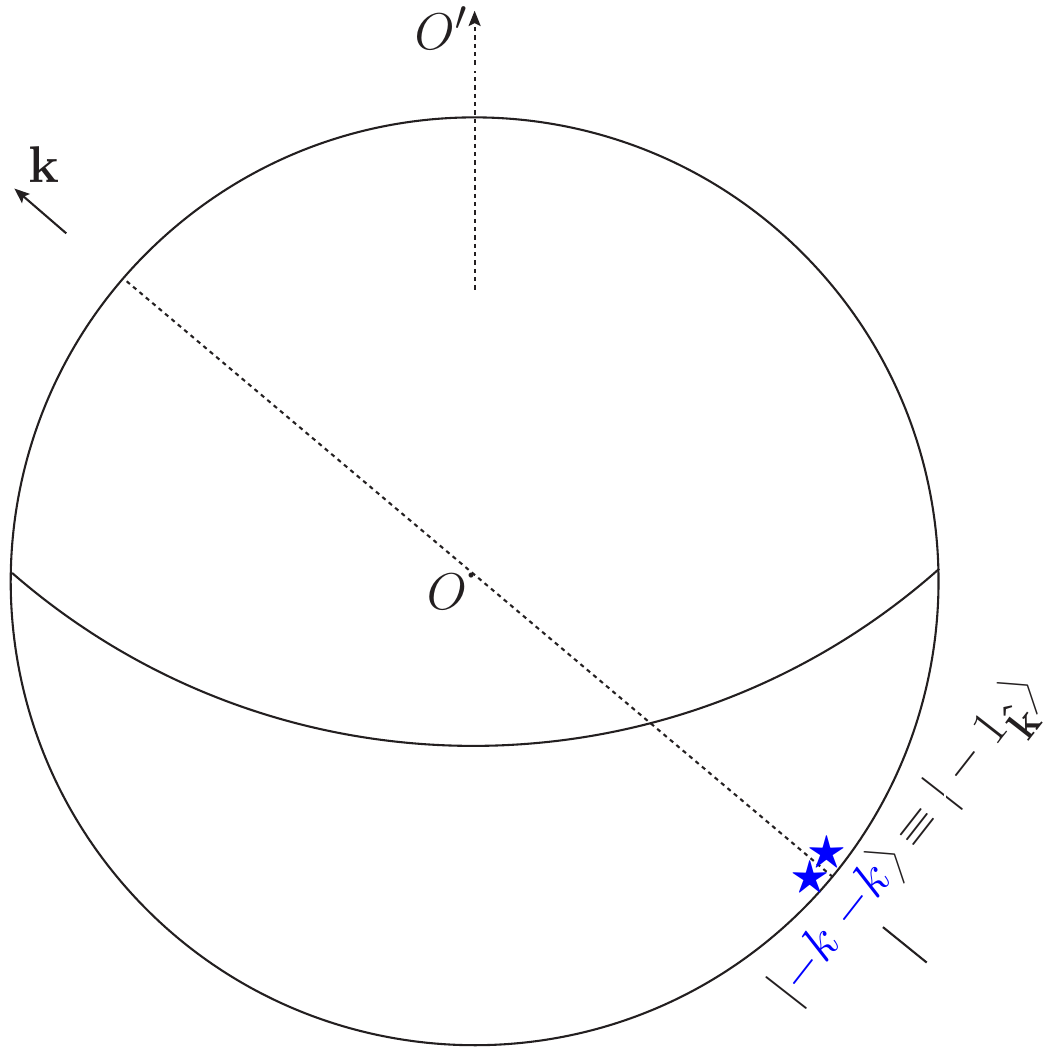}
\caption{$|-1_{\mathbf{k}}\rangle\equiv|{\color{blue}-k-k}\rangle$.}
      \label{fig:MSR3}
  \end{subfigure}
  \caption{\normalsize{Majorana stellar representations of a spin-1 system's eigenstates along a direction $\mathbf k$.}}
  \label{fig:MSR}
\end{figure}

\begin{figure}[h]
\centering\includegraphics[width=0.45\textwidth]{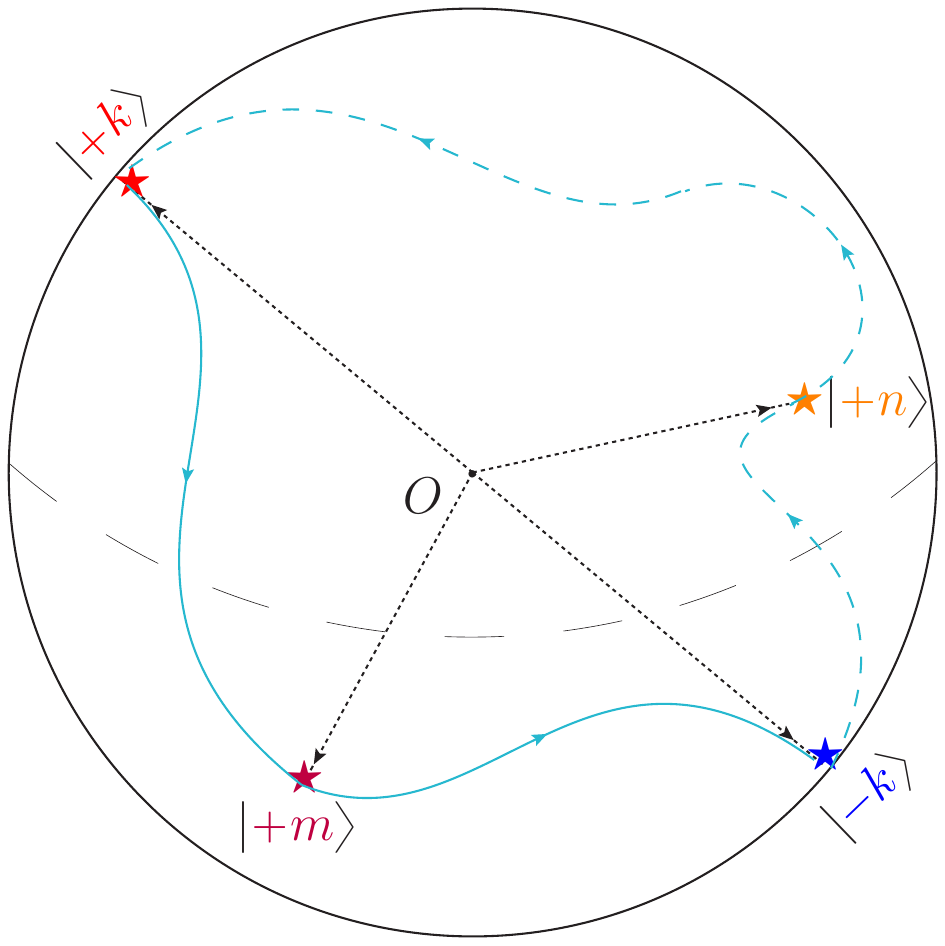}
\caption{\normalsize{A spin-1 state in the Majorana-stellar representation: $|\chi\rangle = \dfrac{1}{\sqrt{3+\mathbf{m}\!\cdot\!\mathbf{n}}}
\big(|{\color{purple}+m}{\color{orange}+n}\rangle
+ |{\color{orange}+n}{\color{purple}+m}\rangle\big)$.
Its projection onto the $s=0$ eigenstate along direction $\mathbf{k}$ is visualized as a trajectory on the BP sphere.}}
 \label{fig:MSRGP}
\end{figure}

\begin{landscape}
\begin{table}
\caption{\label{tab:SmaxHierarchy2}
Maximum--uncertainty sets $S_{\max}^{(n)}$ for $n=1,\dots,5$ KCBS contexts. For each context $G_\alpha=\{A_{\mathbf{k}_{j-1}},A_{\mathbf{k}_{j}},A_{\mathbf{k}_{j+1}}\}$ we denote by $\hat P_{0_{\mathbf{k}_\alpha}} = |0_{\mathbf{k}_\alpha}\rangle\langle 0_{\mathbf{k}_\alpha}|$
the projector onto the $m_s=0$ eigenstate of its central observable. Single--context maximum uncertainty is characterized by $\langle\chi|\hat P_{0_{\mathbf{k}_\alpha}}|\chi\rangle=\tfrac12$.
The multi--context sets $S_{\max}^{(n)}$ are defined as intersections of these level sets. The last column lists the numerically observed upper bounds for the sums of uncertainty products 
$\sum_{\alpha=1}^{n}\Delta A_\alpha\,\Delta C_\alpha$.}
\footnotesize
\begin{tabular}{p{0.6cm} p{2.9cm} p{11.5cm} p{5.9cm}}
\br
$n$ 
& Contexts $G_1,\dots,G_n$
& $n$-context maximum--uncertainty states set $S_{\max}^{(n)}$
& $$\displaystyle \max_{\chi}\sum_{\alpha=1}^{n}\Delta A_\alpha\,\Delta C_\alpha$$
\\
\br

1 
& $G_1=\{A_5,A_1,A_2\}$ 
&
$$S_{\max}^{(1)}[G_1]
=
\left\{
|\chi\rangle\in\mathcal H:\;
\langle\chi|\hat P_{0_{\mathbf{k}_1}}|\chi\rangle=\tfrac12
\right\};
$$
the two--dimensional surface spanned by the axes $\mathbf{\hat{5}}$ and $\mathbf{\hat{2}}$, (see in Fig.~\ref{fig:KCBS_Surface02}).
& 
\[
=1\]
\\[4mm]
\mr
2 
& $G_1=\{A_5,A_1,A_2\}$,
  $G_2=\{A_1,A_2,A_3\}$ 
& $S_{\max}^{(2)}=S_{\max}^{(1)}[G_1] \cap S_{\max}^{(1)}[G_2]$, a one--dimensional curve (the ``orange intersection curve'' in Fig.~\ref{fig:KCBS_Surfaces}).
\[
S_{\max}^{(2)}=
\left\{
|\chi\rangle:\;
\langle\chi|\hat P_{0_{\mathbf{k}_1}}|\chi\rangle=
\langle\chi|\hat P_{0_{\mathbf{k}_2}}|\chi\rangle=\tfrac12
\right\};
\]
numerically 
$|\chi_{\max}^{(2)}\rangle\simeq |0_{\hat{\mathbf u}_2}\rangle$
with a Majorana axis $\hat{\mathbf u}_2$ given by spherical polar coordinates
\[
\theta_{u,2} \approx 0.0326~\mathrm{rad},
\qquad
\varphi_{u,2} \approx 1.8853~\mathrm{rad}.
\]

&
\[
\approx 1.9811
\]
\\[6mm]
\mr
3 
& $G_1,G_2$ above, and
  $G_3=\{A_2,A_3,A_4\}$ 
& $S_{\max}^{(3)}=\bigcap_{\alpha=1}^{3}S_{\max}^{(1)}[G_\alpha]$;
a single pure state, numerically $|\chi_{\max}^{(3)}\rangle\simeq |0_{\hat{\mathbf u}_3}\rangle$ with a Majorana axis 
$\hat{\mathbf u}_3$ given by spherical polar coordinates
\[
\theta_{u,3} \approx 0.0148~\mathrm{rad},
\qquad
\varphi_{u,3} \approx 2.5133~\mathrm{rad}.
\]

&
\[
\approx 2.9681
\]
\\[4mm]
\mr
4 
& $G_1,G_2,G_3$ above, 
  $G_4=\{A_3,A_4,A_5\}$
& $S_{\max}^{(4)}=\bigcap_{\alpha=1}^{4}S_{\max}^{(1)}[G_\alpha]$;
a single pure state, numerically 
$|\chi_{\max}^{(4)}\rangle\simeq |0_{\hat{\mathbf u}_4}\rangle$
with a Majorana axis $\hat{\mathbf u}_4$ given by spherical polar coordinates
\[
\theta_{u,4} \approx 0.0221~\mathrm{rad},
\qquad
\varphi_{u,4} \approx 5.0266~\mathrm{rad}.
\]

&
\[
\approx 3.9592
\]
\\[4mm]
\mr
5 
& All the KCBS contexts:
  $G_1,\dots,G_5$
& $S_{\max}^{(5)} = \bigcap_{\alpha=1}^{5} S_{\max}^{(1)}[G_\alpha]$; the unique KCBS--symmetry axis,
$|\chi_{\max}^{(5)}\rangle = |0_{\hat z}\rangle.$
&
\[
=4(\sqrt5-1)
\approx 4.9443
\]\\[4mm]
\br
\end{tabular}
\end{table}
\end{landscape}

\section*{References}
\bibliographystyle{ieeetr}
\bibliography{References}

\end{document}


\title[Supplementary Information]{Supplementary Information for\\
\textit{Information-Theoretic and Operational Measures of Quantum Contextuality}}

\author{A C G\"unhan$^{1,\ast}$ and Z Gedik$^2$}
\address{$^1$ Department of Physics, Mersin University, \c{C}iftlikk\"oy Merkez Yerle\c{s}kesi, Yeni\c{s}ehir, Mersin, 33150 T\"urkiye}
\address{$^2$ Faculty of Engineering and Natural Sciences, Sabanci University, Tuzla, Istanbul, 34956 T\"urkiye}

\ead{\href{mailto:alicangunhan@mersin.edu.tr}{alicangunhan@mersin.edu.tr}} 
\ead{\href{mailto:gedik@sabanciuniv.edu}{gedik@sabanciuniv.edu}}

\begin{indented}
\item[]$^\ast$ Corresponding author.
\end{indented}

\begin{abstract}
This Supplementary Information collects the detailed mathematical derivations
supporting the main text, including the projection-based formulation of the mutual
information energy, the bounds for the operational contextuality measure, the
KCBS uncertainty optimization, and the proof of the mutually-unbiased trace identity.
\end{abstract}

\maketitle

\tableofcontents

\section*{Notation and conventions}
\addcontentsline{toc}{section}{Notation and conventions}

Throughout this Supplementary Information we follow the notation and conventions
of the main text. In particular, $d$ denotes the Hilbert-space dimension,
$\hat A$, $\hat B$, and $\hat C$ denote observables with spectral projectors
$\{\hat P_i^{A,B}\}$ and $\{\hat P_j^{C,B}\}$, and $E(B;A,C)$ is the mutual information
energy defined in Eq.~(3) of the main article.

\section{Proof of the mutually-unbiased trace identity}
\label{app:MutUnbTrProof}

In this appendix we provide a complete proof of the trace identity
\begin{equation}\label{eq:MutUnbTr}
\mathrm{Tr}\!\left[(\hat P^{A,B}_i \hat P^{C,B}_j)^2\right] 
    = \frac{\mathrm{dim}(\hat P^{A,B}_i)\,\mathrm{dim}(\hat P^{C,B}_j)}{d^2},
\end{equation}
(Eq.~(7) of the main text) to characterize the extremal value of the mutual information energy when the eigenspaces of $\hat A$ and $\hat C$ are mutually unbiased within each eigenspace of $\hat B$. Equation~(\ref{eq:MutUnbTr}) expresses the Hilbert--Schmidt overlap $\mathrm{Tr}[(\hat P_i^{A,B} \hat P_j^{C,B})^2]$ between the joint projectors of two measurement contexts in terms of the dimensions of their
eigenspaces. The result relies on the assumption that the eigenbases of the observables are mutually unbiased within each $\hat{B}$--eigenspace, an idealized symmetry condition under which each rank–1 component contributes uniformly to the overlap. Although Eq.~(\ref{eq:MutUnbTr}) is used in the main text only as a structural identity within the mutual information energy, its proof is provided here for completeness.

Let $\mathcal{H}$ be a $d$--dimensional Hilbert space. 
Consider two observables $\hat A$ and $\hat C$ with (possibly degenerate) spectral projectors $\{\hat P_i\}$ and $\{\hat Q_j\}$. Choose orthonormal eigenbases 
$\{|a_k\rangle\}_{k=1}^d$ and $\{|c_\ell\rangle\}_{\ell=1}^d$ 
of $\hat A$ and $\hat C$, respectively.
For each $i$ and $j$, let $\mathcal{I}_i\subseteq\{1,\dots, d\}$ and $\mathcal{J}_j\subseteq\{1,\dots, d\}$ denote the index sets corresponding to the
eigenspaces of $\hat A$ and $\hat C$, respectively.  Then
\[
\hat P_i = \sum_{k\in \mathcal{I}_i} |a_k\rangle\langle a_k|, 
\qquad
\hat Q_j = \sum_{\ell\in \mathcal{J}_j} |c_\ell\rangle\langle c_\ell|.
\]
We assume that the eigenbases are mutually unbiased, i.e.
\begin{equation}
|\langle a_k|c_\ell\rangle|^2 = \frac{1}{d}
\qquad\forall k,\ell.
\label{eq:MUB_condition}
\end{equation}
Define the rank--$1$ projectors
$\hat P_k := |a_k\rangle\langle a_k|$ and 
$\hat Q_\ell := |c_\ell\rangle\langle c_\ell|$.
Then
\[
\hat P_i \hat Q_j 
= \sum_{k\in\mathcal{I}_i}\sum_{\ell\in\mathcal{J}_j} \hat P_k \hat Q_\ell.
\]
Hence
\[
(\hat P_i \hat Q_j)^2 
= \hat P_i \hat Q_j \hat P_i \hat Q_j
= 
\sum_{k,\ell}\sum_{k',\ell'}
\hat P_k \hat Q_\ell \hat P_{k'} \hat Q_{\ell'},
\]
where $k,k'\in \mathcal{I}_i$ and $\ell,\ell'\in \mathcal{J}_j$.
Using orthogonality of the rank--$1$ projectors,
\[
\hat P_k \hat P_{k'} = \delta_{kk'} \hat P_k,
\qquad
\hat Q_\ell \hat Q_{\ell'} = \delta_{\ell\ell'} \hat Q_\ell,
\]
we obtain
\[
\hat P_k \hat Q_\ell \hat P_{k'} \hat Q_{\ell'}
= \delta_{kk'}\,\delta_{\ell\ell'}\,(\hat P_k \hat Q_\ell)^2.
\]
Therefore
\begin{equation}
(\hat P_i\hat  Q_j)^2 
= 
\sum_{k\in \mathcal{I}_i}\sum_{\ell\in \mathcal{J}_j} (\hat P_k \hat Q_\ell)^2.
\label{eq:PiQi_sq_sum}
\end{equation}
Taking the trace and using~(\ref{eq:PiQi_sq_sum}) yields
\[
\mathrm{Tr}\!\left[(\hat P_i \hat Q_j)^2\right]
=
\sum_{k\in \mathcal{I}_i}\sum_{\ell\in \mathcal{J}_j}
\mathrm{Tr}\!\left[(\hat P_k \hat Q_\ell)^2\right].
\]
For rank--$1$ projectors $\hat P_k=|a_k\rangle\langle a_k|$ and $\hat Q_\ell=|c_\ell\rangle\langle c_\ell|$ we have
\[
\hat P_k \hat Q_\ell 
= 
\langle a_k|c_\ell\rangle\,|a_k\rangle\langle c_\ell|,
\]
hence
\[
\mathrm{Tr}[(\hat P_k \hat Q_\ell)^2]
= 
|\langle a_k|c_\ell\rangle|^4.
\]
By the mutual unbiasedness condition~(\ref{eq:MUB_condition}),
$|\langle a_k|c_\ell\rangle|^2 = 1/d$, and therefore
\[
\mathrm{Tr}[(\hat P_k \hat Q_\ell)^2] = \frac{1}{d^2}.
\]
Substituting into the trace sum,
\[
\mathrm{Tr}[(\hat P_i \hat Q_j)^2]
=
\sum_{k\in \mathcal{I}_i}\sum_{\ell\in \mathcal{J}_j} \frac{1}{d^2}
=
\frac{|\mathcal{J}_i|\,|\mathcal{J}_j|}{d^2}.
\]
Since $|\mathcal{I}_i| = \dim \hat P_i$ and $|\mathcal{J}_j| = \dim \hat Q_j$, we conclude
\[
\mathrm{Tr}\!\left[(\hat P_i \hat Q_j)^2\right]
=
\frac{\dim \hat P_i\;\dim \hat Q_j}{d^2}.
\]

\section{Projection-based correspondence between MIE and projector commutators}\label{app:Proj_MIE}

%

In this appendix we establish an exact correspondence between the mutual information energy (MIE) and the Hilbert--Schmidt norm of projector commutators. This correspondence provides the foundation for the bounds derived in the following section~\ref{app:SpectralBound}.

\subsection*{Setup and notation}

Consider a context $G_\alpha = \{A_\alpha, B_\alpha, C_\alpha\}$ of three observables satisfying $[\hat B_\alpha, \hat A_\alpha] = 0 = [\hat B_\alpha, \hat C_\alpha]$, where $\hat A_\alpha$, $\hat B_\alpha$, and $\hat C_\alpha$ denote the Hermitian operators representing the observables $A_\alpha$, $B_\alpha$, and $C_\alpha$, respectively. We work in a $d$-dimensional Hilbert space $\mathcal{H}$ with $d = \dim(\mathcal{H})$.

Let $\hat P_i \equiv \hat P^{A_\alpha,B_\alpha}_i$ and $\hat Q_j \equiv \hat P^{C_\alpha,B_\alpha}_j$ denote the projectors onto the joint eigenspaces of $(\hat A_\alpha, \hat B_\alpha)$ and $(\hat C_\alpha, \hat B_\alpha)$, respectively. These projectors satisfy the completeness relations $\sum_i \hat P_i = \hat{\mathbb{I}}$ and $\sum_j \hat Q_j = \hat{\mathbb{I}}$.

The basis-independent definition of the mutual information energy is
\begin{equation}
E(B_\alpha; A_\alpha, C_\alpha) = \frac{1}{d} \sum_{i,j} \mathrm{Tr}\bigl[(\hat P_i \hat Q_j)^2\bigr].
\label{eq:E_MIE_app}
\end{equation}

Throughout this work we employ the Hilbert--Schmidt norm $\|\hat X\|_{\mathrm{HS}}^2 = \mathrm{Tr}(\hat X^\dagger \hat X)$.

\subsection*{Projector commutator identity}

For any two Hermitian projectors $\hat P_i$ and $\hat Q_j$ satisfying $\hat P_i^2 = \hat P_i$ and $\hat Q_j^2 = \hat Q_j$, the following identity holds:
\begin{equation}
\|[\hat P_i, \hat Q_j]\|_{\mathrm{HS}}^2 = 2\,\mathrm{Tr}(\hat P_i \hat Q_j) - 2\,\mathrm{Tr}\bigl[(\hat P_i \hat Q_j)^2\bigr].
\label{eq:PQlemma}
\end{equation}

\textit{Proof.}
The commutator of two Hermitian operators is anti-Hermitian: $[\hat P_i, \hat Q_j]^\dagger = -[\hat P_i, \hat Q_j]$. Therefore,
\[
\|[\hat P_i, \hat Q_j]\|_{\mathrm{HS}}^2 = \mathrm{Tr}\bigl([\hat P_i, \hat Q_j]^\dagger [\hat P_i, \hat Q_j]\bigr) = -\mathrm{Tr}\bigl([\hat P_i, \hat Q_j]^2\bigr).
\]
Expanding the square of the commutator,
\[
[\hat P_i, \hat Q_j]^2 = \hat P_i \hat Q_j \hat P_i \hat Q_j - \hat P_i \hat Q_j^2 \hat P_i - \hat Q_j \hat P_i^2 \hat Q_j + \hat Q_j \hat P_i \hat Q_j \hat P_i.
\]
Using the projector properties $\hat P_i^2 = \hat P_i$ and $\hat Q_j^2 = \hat Q_j$, and applying the cyclic property of the trace, we obtain
\[
\mathrm{Tr}\bigl([\hat P_i, \hat Q_j]^2\bigr) = 2\,\mathrm{Tr}\bigl[(\hat P_i \hat Q_j)^2\bigr] - 2\,\mathrm{Tr}(\hat P_i \hat Q_j),
\]
from which the identity~(\ref{eq:PQlemma}) follows.

\subsection*{MIE--commutator correspondence}

Summing~(\ref{eq:PQlemma}) over all pairs $(i,j)$ and using the completeness relations, we find
\[
\sum_{i,j} \mathrm{Tr}(\hat P_i \hat Q_j) = \mathrm{Tr}\biggl(\sum_i \hat P_i \cdot \sum_j \hat Q_j\biggr) = \mathrm{Tr}(\hat{\mathbb{I}}) = d.
\]
Combining this with the definition~(\ref{eq:E_MIE_app}) yields
\begin{equation}
1 - E(B_\alpha; A_\alpha, C_\alpha) = \frac{1}{2d} \sum_{i,j} \|[\hat P_i, \hat Q_j]\|_{\mathrm{HS}}^2.
\label{eq:E_comm_correspondence}
\end{equation}

Equation~(\ref{eq:E_comm_correspondence}) establishes an exact quantitative correspondence between the informational overlap, as measured by the MIE, and the algebraic noncommutativity of the subspace projectors. When all projectors commute, $[\hat P_i, \hat Q_j] = 0$ for all $i,j$, and $E = 1$. Conversely, increasing $\|[\hat P_i, \hat Q_j]\|_{\mathrm{HS}}^2$ reduces $E$, signalling the presence of contextuality.

\section{Spectral and tight bounds for the operational contextuality measure}\label{app:SpectralBound}

In this appendix we derive bounds for the operational contextuality measure. We first establish a spectral bound using the projector--commutator correspondence from section~\ref{app:Proj_MIE}, then refine it by incorporating the purity of the quantum state. Finally, we combine this with an operator norm estimate to obtain the tightest bound available from these methods.

The bounds hold mathematically for any finite-dimensional Hilbert space with $d \ge 1$. However, nontrivial contextuality scenarios require $d \ge 3$ by the Kochen--Specker theorem, since for $d = 2$ (qubit systems) the commutation relations $[\hat B, \hat A] = 0 = [\hat B, \hat C]$ together with $[\hat A, \hat C] \neq 0$ cannot be simultaneously satisfied.

\subsection*{Notation and spectral decompositions}

We work in a $d$-dimensional Hilbert space $\mathcal{H}$. For each context $G_\alpha = \{A_\alpha, B_\alpha, C_\alpha\}$ of observables satisfying $[\hat A_\alpha, \hat B_\alpha] = 0 = [\hat B_\alpha, \hat C_\alpha]$, let
\begin{equation}
\hat A_\alpha = \sum_i a_{\alpha i}\, \hat P_{\alpha i}, \qquad \hat C_\alpha = \sum_j c_{\alpha j}\, \hat Q_{\alpha j},
\end{equation}
be the spectral decompositions, where $\hat P_{\alpha i} \equiv \hat P^{A_\alpha, B_\alpha}_i$ and $\hat Q_{\alpha j} \equiv \hat P^{C_\alpha, B_\alpha}_j$ denote the projectors onto the joint eigenspaces of $(\hat A_\alpha, \hat B_\alpha)$ and $(\hat C_\alpha, \hat B_\alpha)$, respectively.

The operational contribution from context $G_\alpha$ for a state $\hat\rho$ is defined as
\begin{equation}
D(G_\alpha, \hat\rho) = \bigl|\mathrm{Tr}\bigl([\hat A_\alpha, \hat C_\alpha]\,\hat\rho\bigr)\bigr|.
\end{equation}

\subsection*{Commutator expansion}

Since the projectors $\{\hat P_{\alpha i}\}$ and $\{\hat Q_{\alpha j}\}$ resolve the identity, the commutator of the observable operators expands as
\begin{equation}
[\hat A_\alpha, \hat C_\alpha] = \sum_{i,j} a_{\alpha i}\, c_{\alpha j}\, [\hat P_{\alpha i}, \hat Q_{\alpha j}].
\label{eq:ACcomm_spectral}
\end{equation}

\subsection*{Spectral bound}

Substituting~(\ref{eq:ACcomm_spectral}) into the definition of $D(G_\alpha, \hat\rho)$ and applying the triangle inequality yields
\begin{equation}
D(G_\alpha, \hat\rho) \le \sum_{i,j} |a_{\alpha i}|\, |c_{\alpha j}|\, \bigl|\mathrm{Tr}\bigl([\hat P_{\alpha i}, \hat Q_{\alpha j}]\,\hat\rho\bigr)\bigr|.
\end{equation}
The Cauchy--Schwarz inequality in Hilbert--Schmidt space gives
\begin{equation}
\bigl|\mathrm{Tr}\bigl([\hat P_{\alpha i}, \hat Q_{\alpha j}]\,\hat\rho\bigr)\bigr| \le \|[\hat P_{\alpha i}, \hat Q_{\alpha j}]\|_{\mathrm{HS}}\, \|\hat\rho\|_{\mathrm{HS}} \le \|[\hat P_{\alpha i}, \hat Q_{\alpha j}]\|_{\mathrm{HS}},
\end{equation}
where the last inequality uses $\|\hat\rho\|_{\mathrm{HS}}^2 = \mathrm{Tr}(\hat\rho^2) \le 1$. Thus
\begin{equation}
D(G_\alpha, \hat\rho) \le \sum_{i,j} |a_{\alpha i}|\, |c_{\alpha j}|\, \|[\hat P_{\alpha i}, \hat Q_{\alpha j}]\|_{\mathrm{HS}}.
\label{eq:D_single_intermediate}
\end{equation}
Applying the Cauchy--Schwarz inequality to the double sum in~(\ref{eq:D_single_intermediate}),
\begin{equation}
\sum_{i,j} |a_{\alpha i}|\, |c_{\alpha j}|\, \|[\hat P_{\alpha i}, \hat Q_{\alpha j}]\|_{\mathrm{HS}} \le \sqrt{\sum_{i,j} a_{\alpha i}^2\, c_{\alpha j}^2} \; \sqrt{\sum_{i,j} \|[\hat P_{\alpha i}, \hat Q_{\alpha j}]\|_{\mathrm{HS}}^2}.
\end{equation}
The first factor factorizes:
\begin{equation}
\sum_{i,j} a_{\alpha i}^2\, c_{\alpha j}^2 = \biggl(\sum_i a_{\alpha i}^2\biggr) \biggl(\sum_j c_{\alpha j}^2\biggr).
\end{equation}
For the second factor, we invoke the correspondence~(\ref{eq:E_comm_correspondence}) from section~\ref{app:Proj_MIE}:
\begin{equation}
\sum_{i,j} \|[\hat P_{\alpha i}, \hat Q_{\alpha j}]\|_{\mathrm{HS}}^2 = 2d\, \bigl[1 - E(B_\alpha; A_\alpha, C_\alpha)\bigr].
\end{equation}
Combining these results, the contribution from a single context satisfies
\begin{equation}
D(G_\alpha, \hat\rho) \le \kappa(A_\alpha, C_\alpha)\, \bigl[1 - E(B_\alpha; A_\alpha, C_\alpha)\bigr]^{1/2},
\label{eq:D_single_bound}
\end{equation}
where the spectral prefactor is defined as
\begin{equation}
\kappa(A_\alpha, C_\alpha) = \sqrt{2d} \biggl(\sum_i a_{\alpha i}^2\biggr)^{\!1/2} \biggl(\sum_j c_{\alpha j}^2\biggr)^{\!1/2}.
\label{eq:kappa_def}
\end{equation}

\subsection*{Purity-corrected spectral bound}

The bound~(\ref{eq:D_single_bound}) can be tightened by retaining the purity of the state. Defining
\begin{equation}
\beta \equiv \|\hat\rho\|_{\mathrm{HS}}^2 = \mathrm{Tr}(\hat\rho^2),
\end{equation}
which satisfies $1/d \le \beta \le 1$, we have
\begin{equation}
\bigl|\mathrm{Tr}\bigl([\hat P_{\alpha i}, \hat Q_{\alpha j}]\,\hat\rho\bigr)\bigr| \le \sqrt{\beta}\, \|[\hat P_{\alpha i}, \hat Q_{\alpha j}]\|_{\mathrm{HS}}.
\end{equation}
Following the same steps as above yields the purity-corrected bound
\begin{equation}
D(G_\alpha, \hat\rho) \le \sqrt{\beta}\, \kappa(A_\alpha, C_\alpha)\, \bigl[1 - E(B_\alpha; A_\alpha, C_\alpha)\bigr]^{1/2}.
\label{eq:spectral_purity_bound}
\end{equation}
For mixed states with $\beta < 1$, this bound is strictly tighter than~(\ref{eq:D_single_bound}).

\subsection*{Operator norm bound}

An alternative state-independent bound follows from the duality between the operator norm and the trace norm. For any operator $\hat X$ and density matrix $\hat\rho$,
\begin{equation}
\bigl|\mathrm{Tr}(\hat X\, \hat\rho)\bigr| \le \|\hat X\|_{\mathrm{op}}\, \|\hat\rho\|_1 = \|\hat X\|_{\mathrm{op}},
\end{equation}
since $\|\hat\rho\|_1 = \mathrm{Tr}(\hat\rho) = 1$. Applied to the commutator,
\begin{equation}
D(G_\alpha, \hat\rho) \le \|[\hat A_\alpha, \hat C_\alpha]\|_{\mathrm{op}}.
\label{eq:op_norm_bound}
\end{equation}
This bound does not involve the MIE directly, but can be substantially tighter when the observables nearly commute.

\subsection*{Hybrid bound: single context}

The bounds~(\ref{eq:spectral_purity_bound}) and~(\ref{eq:op_norm_bound}) are complementary: the spectral bound captures the role of the MIE and state purity, while the operator norm bound is state-independent and often tight when the commutator is small. Taking the minimum yields the tightest single-context estimate:
\begin{equation}
D(G_\alpha, \hat\rho) \le \min\Bigl\{ \|[\hat A_\alpha, \hat C_\alpha]\|_{\mathrm{op}},\; \sqrt{\beta}\, \kappa(A_\alpha, C_\alpha)\, \bigl[1 - E(B_\alpha; A_\alpha, C_\alpha)\bigr]^{1/2} \Bigr\}.
\label{eq:hybrid_single}
\end{equation}

\subsection*{Global bounds}

For a collection of $N$ contexts $G = \{G_\alpha\}_{\alpha=1}^{N}$, the global bounds are obtained by summing the single-context bounds. The global spectral bound reads
\begin{equation}
D(G, \hat\rho) = \sum_{\alpha=1}^{N} D(G_\alpha, \hat\rho) \le \sum_{\alpha=1}^{N} \kappa(A_\alpha, C_\alpha)\, \bigl[1 - E(B_\alpha; A_\alpha, C_\alpha)\bigr]^{1/2},
\label{eq:Op_Global_Bound}
\end{equation}
and the global hybrid bound is
\begin{equation}
D(G, \hat\rho) \le \sum_{\alpha=1}^{N} \min\Bigl\{ \|[\hat A_\alpha, \hat C_\alpha]\|_{\mathrm{op}},\; \sqrt{\beta}\, \kappa_\alpha\, \bigl[1 - E_\alpha\bigr]^{1/2} \Bigr\},
\label{eq:hybrid_global}
\end{equation}
where $\kappa_\alpha \equiv \kappa(A_\alpha, C_\alpha)$ and $E_\alpha \equiv E(B_\alpha; A_\alpha, C_\alpha)$.

These global bounds are mathematically valid for any collection of contexts. However, when contexts share observables—as occurs in the KCBS scenario where each observable appears in two adjacent contexts—the bounds may be looser than if the contexts were independent. The summation treats each context separately, without accounting for correlations introduced by shared observables. Nevertheless, the bounds remain correct upper limits and provide useful estimates of the total operational contextuality.

\subsection*{Saturation conditions}

The hybrid bound~(\ref{eq:hybrid_single}) is saturated when:
\begin{itemize}
\item For the operator norm branch: $\hat\rho$ is a pure state whose support lies in the eigenspace of $[\hat A_\alpha, \hat C_\alpha]$ corresponding to the eigenvalue of largest magnitude.
\item For the spectral branch: all intermediate Cauchy--Schwarz inequalities are saturated, which requires specific alignment conditions between the state and the projector structure.
\end{itemize}

This completes the derivation of the bounds for the operational contextuality measure.


\section{KCBS observables and uncertainty optimization in the Majorana–stellar representation}\label{app:KCBSUncertainty}

This appendix provides the detailed geometric formalism underlying the Majorana--stellar representation used in Section 5 of the main text. We develop the orthonormal triad decomposition for spin-1 states, derive explicit coefficient formulas, and establish the geometric conditions for maximum uncertainty.

A key advantage of this formalism is computational: once a spin-1 state is expressed in terms of its Majorana directions $\mathbf{m}$ and $\mathbf{n}$, all relevant quantities---expectation values, overlaps, uncertainties---reduce to elementary trigonometric functions of the angles between directions. This bypasses explicit matrix manipulations and renders the geometric content of quantum-mechanical expressions transparent.

\subsection{Spin-1 states as symmetric two-qubit states}
\label{app:MSR_symmetric}

The Majorana--stellar representation~\cite{MajoranaSR1932} maps a spin-$s$ pure state to a constellation on the Bloch sphere. Each basis vector of the $(2s+1)$-dimensional Hilbert space can be represented by a symmetric constellation of $2s$ Majorana stars. Across the full set of $(2s+1)$ eigenstates, one obtains two single-star constellations for $s=\tfrac{1}{2}$, three two-star constellations for $s=1$, four three-star constellations for $s=\tfrac{3}{2}$, and so on. While for $s \le \tfrac{3}{2}$ the full orthogonal set of eigenstates can still be visualized by distinct star constellations on the Bloch--Poincar\'e sphere, for $s \ge 2$ the higher-dimensional orthogonality cannot be faithfully embedded on a two-dimensional surface. Even in those cases, the Majorana--stellar representation remains computationally advantageous and conceptually clarifying.

For spin-1 ($s=1$), the three basis states $|{+}1_{\mathbf{k}}\rangle$, $|0_{\mathbf{k}}\rangle$, $|{-}1_{\mathbf{k}}\rangle$ are each represented by a pair of Majorana stars, and any pure state is a superposition of these basis vectors. The identification proceeds via the symmetric subspace of two spin-$\tfrac{1}{2}$ particles.

Let $|{+k}\rangle$ and $|{-k}\rangle$ denote the spin-up and spin-down states along direction $\mathbf{k}$. The three-dimensional symmetric subspace of $\mathbb{C}^2 \times \mathbb{C}^2$ is spanned by
\begin{equation}\label{eq:symmetric_basis}
|{+k}{+k}\rangle, \quad
\frac{1}{\sqrt{2}}\bigl(|{+k}{-k}\rangle + |{-k}{+k}\rangle\bigr), \quad
|{-k}{-k}\rangle,
\end{equation}
which correspond to the $m_s = +1, 0, -1$ eigenstates of $\hat{S}_{\mathbf{k}}$, respectively. Though the product space has dimension four, restriction to the symmetric subspace yields an effective three-dimensional Hilbert space isomorphic to that of a spin-1 particle.

A general pure spin-1 state with Majorana directions $\mathbf{m}$ and $\mathbf{n}$ on the Bloch sphere takes the form
\begin{equation}\label{eq:general_state_app}
|\chi\rangle = \frac{1}{\sqrt{3 + \mathbf{m}\cdot\mathbf{n}}}\bigl(|{+m}{+n}\rangle + |{+n}{+m}\rangle\bigr),
\end{equation}
where the normalization factor accounts for the overlap between the two spin-$\tfrac{1}{2}$ states. The unit vectors are parametrized by spherical coordinates:
\begin{equation}
\mathbf{m} = (\sin\vartheta_m\cos\varphi_m,\, \sin\vartheta_m\sin\varphi_m,\, \cos\vartheta_m),
\end{equation}
and similarly for $\mathbf{n}$.

\subsection{Orthonormal triad decomposition}
\label{app:MSR_triad}

A particularly useful representation expands $|\chi\rangle$ in terms of an orthonormal basis of $m_s=0$ states along three mutually perpendicular axes. Given a primary direction $\mathbf{k}$, we construct the orthonormal triad $\{\mathbf{k}, \mathbf{k}_1, \mathbf{k}_2\}$ as
\begin{eqnarray}
\mathbf{k}   &=& (\sin\vartheta_k\cos\varphi_k,\; \sin\vartheta_k\sin\varphi_k,\; \cos\vartheta_k), \label{eq:k_def}\\
\mathbf{k}_1 &=& (\cos\vartheta_k\cos\varphi_k,\; \cos\vartheta_k\sin\varphi_k,\; -\sin\vartheta_k), \label{eq:k1_def}\\
\mathbf{k}_2 &=& (-\sin\varphi_k,\; \cos\varphi_k,\; 0). \label{eq:k2_def}
\end{eqnarray}
The corresponding $m_s=0$ states $|\hat{K}\rangle$, $|\hat{K}_1\rangle$, $|\hat{K}_2\rangle$ form an orthonormal basis for the spin-1 Hilbert space, where
\begin{equation}
|\hat{K}\rangle \equiv |0_{\mathbf{k}}\rangle = \frac{1}{\sqrt{2}}\bigl(|{+k}{-k}\rangle + |{-k}{+k}\rangle\bigr),
\end{equation}
and similarly for $|\hat{K}_1\rangle$ and $|\hat{K}_2\rangle$.

Any pure spin-1 state can thus be decomposed as
\begin{equation}\label{eq:triad_decomposition}
|\chi\rangle = K\,|\hat{K}\rangle + K_1\,|\hat{K}_1\rangle + K_2\,|\hat{K}_2\rangle,
\end{equation}
with $|K|^2 + |K_1|^2 + |K_2|^2 = 1$.

\subsection{Explicit coefficient formulas}
\label{app:MSR_coefficients}

For a state $|\chi\rangle$ with Majorana directions $\mathbf{m}$ and $\mathbf{n}$, the coefficients in the triad decomposition~(\ref{eq:triad_decomposition}) are given by
\begin{eqnarray}
K &=& \frac{1}{\mathcal{N}}
\biggl\{
\sin\vartheta_k
\Bigl(\cos\tfrac{\vartheta_m}{2}\cos\tfrac{\vartheta_n}{2}
-\sin\tfrac{\vartheta_m}{2}\sin\tfrac{\vartheta_n}{2}
\,e^{i(\varphi_m+\varphi_n-2\varphi_k)}\Bigr)
\nonumber\\
&&\quad
-\cos\vartheta_k
\Bigl(\sin\tfrac{\vartheta_m}{2}\cos\tfrac{\vartheta_n}{2}
\,e^{i(\varphi_m-\varphi_k)}
+\cos\tfrac{\vartheta_m}{2}\sin\tfrac{\vartheta_n}{2}
\,e^{i(\varphi_n-\varphi_k)}\Bigr)
\biggr\}, \label{eq:K_formula}
\end{eqnarray}
\begin{eqnarray}
K_1 &=& \frac{-1}{\mathcal{N}}
\biggl\{
\cos\vartheta_k
\Bigl(\cos\tfrac{\vartheta_m}{2}\cos\tfrac{\vartheta_n}{2}
-\sin\tfrac{\vartheta_m}{2}\sin\tfrac{\vartheta_n}{2}
\,e^{i(\varphi_m+\varphi_n-2\varphi_k)}\Bigr)
\nonumber\\
&&\quad
+\sin\vartheta_k
\Bigl(\sin\tfrac{\vartheta_m}{2}\cos\tfrac{\vartheta_n}{2}
\,e^{i(\varphi_m-\varphi_k)}
+\cos\tfrac{\vartheta_m}{2}\sin\tfrac{\vartheta_n}{2}
\,e^{i(\varphi_n-\varphi_k)}\Bigr)
\biggr\}, \label{eq:K1_formula}
\end{eqnarray}
\begin{eqnarray}
K_2 &=& \frac{1}{\mathcal{N}}
\Bigl(\cos\tfrac{\vartheta_m}{2}\cos\tfrac{\vartheta_n}{2}
+\sin\tfrac{\vartheta_m}{2}\sin\tfrac{\vartheta_n}{2}
\,e^{i(\varphi_m+\varphi_n-2\varphi_k)}\Bigr), \label{eq:K2_formula}
\end{eqnarray}
where the normalization factor is
\begin{equation}\label{eq:normalization}
\mathcal{N} = \sqrt{1 + \mathbf{m}\cdot\mathbf{n} - \cos\vartheta_m\cos\vartheta_n}.
\end{equation}

These expressions remain well-defined throughout the parameter space except when the two Majorana directions become antipodal ($\mathbf{n} = -\mathbf{m}$). In this degenerate case, the state reduces to a coherent spin state $|{+m}{+m}\rangle$ or $|{-m}{-m}\rangle$. The apparent singularities are removable, and all physical quantities remain finite.

\subsection{Maximum uncertainty condition}
\label{app:MSR_maxuncert}

The expectation value of the projector $\hat{P}_{0_{\mathbf{k}}} = |\hat{K}\rangle\langle\hat{K}|$ in state $|\chi\rangle$ is simply
\begin{equation}\label{eq:projector_expectation}
\langle\chi|\hat{P}_{0_{\mathbf{k}}}|\chi\rangle = |K|^2.
\end{equation}
Since the variance of the dichotomic observable $\hat{A}_{\mathbf{k}} = \hat{I} - 2\hat{P}_{0_{\mathbf{k}}}$ is $(\Delta A_{\mathbf{k}})^2 = 4|K|^2(1-|K|^2)$, maximum uncertainty occurs when
\begin{equation}\label{eq:max_uncert_condition}
|K|^2 = \frac{1}{2}.
\end{equation}
Geometrically, this condition defines a quadratic surface in the coefficient space $(K, K_1, K_2)$. Combined with the normalization constraint $|K|^2 + |K_1|^2 + |K_2|^2 = 1$, the maximum-uncertainty states satisfy $|K_1|^2 + |K_2|^2 = \tfrac{1}{2}$.

For the three-dimensional Euclidean-like space arising from the spin-1 Majorana--stellar representation, this condition manifests as the plane perpendicular to $\mathbf{k}$, spanned by $\mathbf{k}_1$ and $\mathbf{k}_2$. Hence, the states that maximize the uncertainty of $\hat{A}_{\mathbf{k}}$ \textit{lie on this plane}, which makes the MSR a natural tool to both compute and visualize the uncertainty of $\hat{A}_{\mathbf{k}}$ within any given context.

\subsection{Overlap formula and Bargmann invariants}
\label{app:MSR_overlap}

The overlap between a general state $|\chi\rangle$ and the $m_s=0$ state $|\hat{K}\rangle$ can be expressed in terms of angles and geometric phases. From~(\ref{eq:general_state_app}) and~(\ref{eq:triad_decomposition}), one obtains
\begin{eqnarray}
|\langle\hat{K}|\chi\rangle|^2 &=& \frac{2}{3+\mathbf{m}\cdot\mathbf{n}}
\biggl[
\cos^2\!\frac{\gamma_{km}}{2}\sin^2\!\frac{\gamma_{kn}}{2}
+ \cos^2\!\frac{\gamma_{kn}}{2}\sin^2\!\frac{\gamma_{km}}{2}
\nonumber\\
&&\qquad\quad
+ B(+k,+m,-k,+n) + B(+k,+n,-k,+m)
\biggr], \label{eq:overlap_formula}
\end{eqnarray}
where $\gamma_{km} = \angle(\mathbf{k}, \mathbf{m})$ denotes the angle between directions $\mathbf{k}$ and $\mathbf{m}$, and
\begin{equation}\label{eq:Bargmann_def}
B(+k,+m,-k,+n) = \langle{+k}|{+m}\rangle\langle{+m}|{-k}\rangle\langle{-k}|{+n}\rangle\langle{+n}|{+k}\rangle
\end{equation}
is a Bargmann invariant~\cite{Bargmann1964,Berry1984,Sjoqvist2006,Gedik2021,LiuFu2014}. This four-vertex invariant equals the geometric phase acquired along the closed path $|{+k}\rangle \to |{+m}\rangle \to |{-k}\rangle \to |{+n}\rangle \to |{+k}\rangle$ on the Bloch sphere.

The appearance of Bargmann invariants in~(\ref{eq:overlap_formula}) reflects the intrinsically geometric nature of the maximum-uncertainty condition: determining whether a state lies on a maximum-uncertainty surface involves not only the pairwise angles between directions but also the oriented area enclosed by the associated spherical polygon.

\subsection{Application to KCBS contexts}
\label{app:MSR_KCBS}

For the KCBS scenario, the five directions $\mathbf{\hat 1}, \ldots, \mathbf{\hat 5}$ define five maximum-uncertainty planes in the three-dimensional Euclidean-like space. A state $|\chi\rangle$ achieves maximum uncertainty for the observable $\hat{A}_{\mathbf{k}}$ if and only if its coefficient $K$ (in the triad decomposition centered on $\mathbf{k}$) satisfies $|K|^2 = \tfrac{1}{2}$.

For a single context, this condition defines a two-dimensional surface. For two contexts, the intersection is generically a one-dimensional curve. For three or more contexts, the corresponding planes do not generically share a common line; hence true maximum uncertainty cannot be achieved simultaneously for all contexts. For the full KCBS family of five contexts, the optimization yields a unique state: $|0_{\mathbf{z}}\rangle$, the $m_s=0$ eigenstate along the pentagonal symmetry axis (see Table~\ref{tab:SmaxHierarchy2}).

This analysis provides a geometric proof that $|0_{\mathbf{z}}\rangle$ uniquely achieves the optimal sum of uncertainty products $\sum_{\alpha=1}^{5}(\Delta A_{\alpha-1})(\Delta A_{\alpha+1})$ over all pure spin-1 states.



\section*{}
\addcontentsline{toc}{section}{E Maximum--uncertainty sets $S_{\max}^{(n)}$ for $n=1,\dots,5$ KCBS contexts.}
\begin{landscape}
\begin{table}
\caption{\label{tab:SmaxHierarchy2}
Maximum--uncertainty sets $S_{\max}^{(n)}$ for $n=1,\dots,5$ KCBS contexts. For each context $G_\alpha=\{A_{\mathbf{k}_{j-1}},A_{\mathbf{k}_{j}},A_{\mathbf{k}_{j+1}}\}$ we denote by $\hat P_{0_{\mathbf{k}_\alpha}} = |0_{\mathbf{k}_\alpha}\rangle\langle 0_{\mathbf{k}_\alpha}|$
the projector onto the $m_s=0$ eigenstate of its central observable. Single--context maximum uncertainty is characterized by $\langle\chi|\hat P_{0_{\mathbf{k}_\alpha}}|\chi\rangle=\tfrac12$.
The multi--context sets $S_{\max}^{(n)}$ are defined as intersections of these level sets. The last column lists the numerically observed upper bounds for the sums of uncertainty products 
$\sum_{\alpha=1}^{n}\Delta A_\alpha\,\Delta C_\alpha$.}
\footnotesize
\begin{tabular}{p{0.6cm} p{2.9cm} p{11.5cm} p{5.9cm}}
\br
$n$ 
& Contexts $G_1,\dots,G_n$
& $n$-context maximum--uncertainty states set $S_{\max}^{(n)}$
& $$\displaystyle \max_{\chi}\sum_{\alpha=1}^{n}\Delta A_\alpha\,\Delta C_\alpha$$
\\
\br

1 
& $G_1=\{A_5,A_1,A_2\}$ 
&
$$S_{\max}^{(1)}[G_1]
=
\left\{
|\chi\rangle\in\mathcal H:\;
\langle\chi|\hat P_{0_{\mathbf{k}_1}}|\chi\rangle=\tfrac12
\right\};
$$
the two--dimensional surface spanned by the axes $\mathbf{\hat{5}}$ and $\mathbf{\hat{2}}$, (see in Fig. 4c).
& 
\[
=1\]
\\[4mm]
\mr
2 
& $G_1=\{A_5,A_1,A_2\}$,
  $G_2=\{A_1,A_2,A_3\}$ 
& $S_{\max}^{(2)}=S_{\max}^{(1)}[G_1] \cap S_{\max}^{(1)}[G_2]$, a one--dimensional curve (the ``orange curve'' in Fig. 4b).
\[
S_{\max}^{(2)}=
\left\{
|\chi\rangle:\;
\langle\chi|\hat P_{0_{\mathbf{k}_1}}|\chi\rangle=
\langle\chi|\hat P_{0_{\mathbf{k}_2}}|\chi\rangle=\tfrac12
\right\};
\]
numerically 
$|\chi_{\max}^{(2)}\rangle\simeq |0_{\hat{\mathbf u}_2}\rangle$
with a Majorana axis $\hat{\mathbf u}_2$ given by spherical polar coordinates
\[
\theta_{u,2} \approx 0.0326~\mathrm{rad},
\qquad
\varphi_{u,2} \approx 1.8853~\mathrm{rad}.
\]

&
\[
\approx 1.9811
\]
\\[6mm]
\mr
3 
& $G_1,G_2$ above, and
  $G_3=\{A_2,A_3,A_4\}$ 
& $S_{\max}^{(3)}=\bigcap_{\alpha=1}^{3}S_{\max}^{(1)}[G_\alpha]$;
a single pure state, numerically $|\chi_{\max}^{(3)}\rangle\simeq |0_{\hat{\mathbf u}_3}\rangle$ with a Majorana axis 
$\hat{\mathbf u}_3$ given by spherical polar coordinates
\[
\theta_{u,3} \approx 0.0148~\mathrm{rad},
\qquad
\varphi_{u,3} \approx 2.5133~\mathrm{rad}.
\]

&
\[
\approx 2.9681
\]
\\[4mm]
\mr
4 
& $G_1,G_2,G_3$ above, 
  $G_4=\{A_3,A_4,A_5\}$
& $S_{\max}^{(4)}=\bigcap_{\alpha=1}^{4}S_{\max}^{(1)}[G_\alpha]$;
a single pure state, numerically 
$|\chi_{\max}^{(4)}\rangle\simeq |0_{\hat{\mathbf u}_4}\rangle$
with a Majorana axis $\hat{\mathbf u}_4$ given by spherical polar coordinates
\[
\theta_{u,4} \approx 0.0221~\mathrm{rad},
\qquad
\varphi_{u,4} \approx 5.0266~\mathrm{rad}.
\]

&
\[
\approx 3.9592
\]
\\[4mm]
\mr
5 
& All the KCBS contexts:
  $G_1,\dots,G_5$
& $S_{\max}^{(5)} = \bigcap_{\alpha=1}^{5} S_{\max}^{(1)}[G_\alpha]$; the unique KCBS--symmetry axis,
$|\chi_{\max}^{(5)}\rangle = |0_{\hat z}\rangle.$
&
\[
=4(\sqrt5-1)
\approx 4.9443
\]\\[4mm]
\br
\end{tabular}
\end{table}
\end{landscape}

\section*{References}
\bibliographystyle{ieeetr}
\bibliography{References}